\documentclass[aps,prd,twocolumn,superscriptaddress,showpacs,amsmath,amssymb]{revtex4-2}

\usepackage[caption=false]{subfig}
\usepackage{upgreek}
\usepackage{standalone}
\usepackage{booktabs}
\usepackage{enumerate}
\usepackage{float}

\usepackage{graphicx} %
\usepackage{dcolumn} %
\usepackage{xcolor} %
\usepackage{amsmath,amssymb} %
\usepackage{hyperref} %
\usepackage[utf8]{inputenc} %
\usepackage[english]{babel} %
\usepackage[displaymath, mathlines]{lineno} %
\usepackage{blindtext} %
\usepackage{ifthen} %
\usepackage{orcidlink} %
\usepackage{cancel}

\makeatletter\AtBeginDocument{\let\@elt\relax}\makeatother

\graphicspath{{figures/}} %

\newboolean{articletitles}
\setboolean{articletitles}{true} %

\addto\captionsenglish{}
\addto\captionsenglish{}

\input{belle2-symbols}
\newcommand{\BKnn}{\ensuremath{B^{+}\to K^{+}\nu\bar{\nu}}\xspace}
\newcommand{\BJpsiK}{\ensuremath{B^{+}\to K^{+} \jpsi}\xspace}
\def\tomumu{\ensuremath{\to\mu^+\mu^-}}

\def\Jpsislash{\cancel{\jpsi}}

\newcommand{\BJpsiKSlash}{\ensuremath{B^{+}\to K^{+} {\Jpsislash}\xspace}}

\renewcommand{\pt}{\ensuremath{p_\mathrm{T}}\xspace}

\newcommand{\dz}{\ensuremath{d_{z}}\xspace}
\newcommand{\dr}{\ensuremath{d_{r}}\xspace}
\newcommand{\sumEp}{\ensuremath{E^*_\mathrm{miss}+cp^*_\mathrm{miss}}\xspace}
\newcommand{\Eecl}{\ensuremath{E_\mathrm{extra}}\xspace}
\newcommand{\nge}{\ensuremath{n_{\gamma\,\mathrm{extra}}}\xspace}
\newcommand{\Btag}{\ensuremath{B_\mathrm{tag}}\xspace}
\def\BDT#1{\ensuremath{\mathrm{BDT}_#1}\xspace}

\def\combinationBF{\ensuremath{(2.3 \pm 0.7)\times 10^{-5}}\xspace}

\def\combinationBFdetailed{\ensuremath{\left[2.3 \pm 0.5(\mathrm{stat})^{+0.5}_{-0.4}(\mathrm{syst})\right]\times 10^{-5}}\xspace}

\def\combinationsigSM{\ensuremath{2.7}\xspace}

\def\combinationsigO{\ensuremath{3.5}\xspace}

\def\ITABF{\ensuremath{(2.7\pm 0.7)\times 10^{-5}}\xspace}

\def\ITABFdetailed{\ensuremath{\left[2.7\pm 0.5(\mathrm{stat})\pm 0.5(\mathrm{syst})\right] \times 10^{-5}}\xspace}

\def\ITAsigO{\ensuremath{3.5}\xspace}

\def\ITAsigSM{\ensuremath{2.9}\xspace}

\def\HTABF{\ensuremath{(1.1^{+1.2}_{-1.0})\times 10^{-5}}\xspace}

\def\HTABFdetailed{\ensuremath{\left[1.1^{+0.9}_{-0.8}(\mathrm{stat}){}^{+0.8}_{-0.5}(\mathrm{syst})\right] \times 10^{-5}}\xspace}

\def\HTAsigO{\ensuremath{1.1}\xspace}

\def\HTAsigSM{\ensuremath{0.6}\xspace}
\usepackage[capitalise]{cleveref}
\crefname{section}{Sec.}{Secs.}
\crefname{table}{Table}{Tabs.}

\begin{document}

\title{Evidence for $\boldsymbol{\BKnn}$ decays}
  \author{I.~Adachi\,\orcidlink{0000-0003-2287-0173}} %
  \author{K.~Adamczyk\,\orcidlink{0000-0001-6208-0876}} %
  \author{L.~Aggarwal\,\orcidlink{0000-0002-0909-7537}} %
  \author{H.~Ahmed\,\orcidlink{0000-0003-3976-7498}} %
  \author{H.~Aihara\,\orcidlink{0000-0002-1907-5964}} %
  \author{N.~Akopov\,\orcidlink{0000-0002-4425-2096}} %
  \author{A.~Aloisio\,\orcidlink{0000-0002-3883-6693}} %
  \author{N.~Anh~Ky\,\orcidlink{0000-0003-0471-197X}} %
  \author{D.~M.~Asner\,\orcidlink{0000-0002-1586-5790}} %
  \author{H.~Atmacan\,\orcidlink{0000-0003-2435-501X}} %
% ...
%
\collaboration{The Belle II Collaboration}

\begin{abstract}
We search for the rare decay \BKnn in a $362\invfb$ sample of electron-positron collisions at the \Y4S resonance collected with the \belletwo detector at the SuperKEKB collider.
We use the inclusive properties of the accompanying \B meson in $\Y4S \to \BBbar$ events
to suppress background from other decays of the signal \B candidate and light-quark pair production.
We validate the measurement with an auxiliary analysis based on a conventional hadronic reconstruction of the accompanying $\B$ meson.
For background suppression, we exploit distinct signal features using 
machine learning methods tuned with simulated data. The signal-reconstruction efficiency and background suppression are validated through various control channels. The branching fraction is extracted in a maximum likelihood fit. Our inclusive and hadronic analyses 
yield consistent results for the 
$B^{+}\rightarrow K^{+}\nu\bar{\nu}$ 
branching fraction
of \ITABFdetailed and \HTABFdetailed, respectively.
Combining the results, 
we determine the branching fraction of the decay $B^{+}\rightarrow K^{+}\nu\bar{\nu}$ to be \combinationBFdetailed, providing the first evidence for this decay at \combinationsigO standard deviations.
The combined result is \combinationsigSM standard deviations above the standard model expectation.

\end{abstract}

\maketitle

\section{Introduction}

Flavor-changing neutral-current transitions, such as $b\to s \nu\bar{\nu}$ and $b\to s \ell\ell$, where $\ell$ represents a charged lepton, are suppressed in the standard model (SM) of particle physics, because of the Glashow-Iliopoulos-Maiani mechanism \cite{Glashow:1970gm}. These transitions can only occur at higher orders in SM perturbation theory through weak-interaction amplitudes that involve the exchange of at least two gauge bosons. 
Rate predictions for $b \to s\ell\ell$ have significant
theoretical uncertainties from the breakdown of factorization due to
photon exchange~\cite{Buras:2014fpa}. This process does not contribute to $b \to
s \nu \bar{\nu}$, so the corresponding rate predictions are relatively
precise.

\begin{figure}[hptb]
\centering
\begin{tabular}{cc}
{\includegraphics[width=0.49\linewidth]{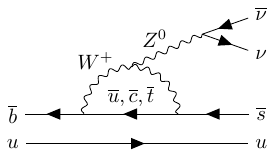}} &
{\includegraphics[width=0.49\linewidth]{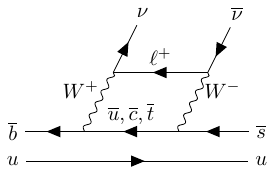}} \\
a) & b)\\
\end{tabular}
\begin{tabular}{c}
{\includegraphics[width=.60\linewidth]{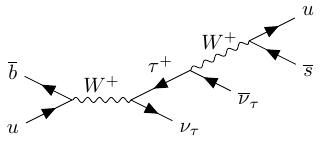}}\\
c)
\end{tabular}
\caption{Lowest-order quark-level diagrams for the \BKnn decay in the SM are either of the penguin (a), or box type (b): examples are shown. The long-distance double-charged-current diagram (c) arising at tree level in the SM also contributes to the \BKnn decay.}
\label{fig:feynman}
\end{figure}

The $b\to s \nu\bar{\nu}$ transition provides the leading amplitudes for the 
\BKnn decay in the SM, as shown in \cref{fig:feynman}.
The SM branching fraction of the \BKnn decay \cite{chargeConj} 
is predicted in Ref.~\cite{Parrott:2022zte} to be
\begin{equation}
\mathcal{B}(\BKnn) =
\left(5.58 \pm 0.37 \right)\times 10^{-6}\,,
\label{eq:brf}
\end{equation}
including a contribution of $\left(0.61 \pm 0.06\right)\times 10^{-6}$ from the long-distance double-charged-current $B^+\to \tau^+(\to K^+\bar{\nu})\nu$ decay.
The \BKnn decay rate can be significantly modified
in models that predict non-SM particles, such as leptoquarks \cite{Becirevic:2018afm}.
In addition, the $B^+$ meson could decay into a kaon and an undetectable particle, such as an axion \cite{MartinCamalich:2020dfe} or a dark-sector mediator \cite{Filimonova:2019tuy}.

In all analyses reported to date \cite{PhysRevLett.86.2950,PhysRevD.87.111103,PhysRevD.87.112005,PhysRevD.82.112002,PhysRevD.96.091101,Belle-II:2021rof}, no evidence for a signal has been found, and the current experimental upper limit on the branching fraction is $1.6\times10^{-5}$ at the 90\% confidence level \cite{ParticleDataGroup:2022pth}.
The study of the \BKnn decay is experimentally challenging as the final state contains two neutrinos that are not reconstructed.
This prevents the full reconstruction of the kinematic properties of the decay, hindering the differentiation of signal distributions from background.

In this study the signal $B$ meson is produced in the $e^+ e^- \to \Y4S \to B^+ B^-$ process. 
The at-threshold production of $\BBbar$ pairs helps to mitigate the limitations due to the unconstrained kinematics, as the partner $B$ meson can be used to infer the presence and properties of the signal $B$. 
An inclusive tagging analysis method (ITA) exploiting inclusive properties from the \B meson pair-produced along with the signal $B$, is applied to the entire Belle II data set currently available, superseding the results of Ref.~\cite{Belle-II:2021rof}, where this method was first used. 
In addition, an auxiliary analysis using the well-established hadronic tagging analysis method (HTA) \cite{PhysRevD.87.111103,PhysRevD.87.112005} is presented;  this involves explicit reconstruction of the partner \B meson through a hadronic decay. 
The HTA method offers an important consistency check of the newer inclusive tagging method and helps validate the ITA results. In addition, the small size of the overlap between the HTA and ITA samples allows for  a straightforward 
 combination of the results, achieving a 10\% increase in precision over the ITA result alone. 

The ITA commences with the reconstruction of charged and neutral particles, followed by 
the selection of a single signal kaon candidate in events with one or more kaons.
Subsequently, relevant quantities are computed using the kaon candidate, along with the remaining particles in the event, to discriminate between signal and background processes. These quantities are used in boosted decision trees (BDTs) \cite{FastBDTBelleII,Chen:2016:XST:2939672.2939785} that are optimized and trained using simulated data. A signal region is then defined, and a binned profile-likelihood sample-composition fit is carried out on data. This fit uses simulated samples to provide predictions to determine the branching fraction of the \BKnn decay along with the rates of background processes. The fit incorporates systematic uncertainties arising from detector and physics-modeling imperfections as nuisance parameters. To validate the modeling of signal and background processes in simulation, several control channels are employed. The method is further validated through a closure-test measurement of the branching fraction of the $B^+ \to \pi^+K^0$ decay.

The HTA follows a similar method, but begins with the reconstruction of the partner \B meson, and then proceeds
to the definition of the signal candidate.

Except for the tagging method, the two analyses are similar in terms of particle reconstruction, event selection, usage of control samples,
fit strategy, and treatment of common systematic uncertainties. In what follows, common approaches and details of the ITA are given first, followed by the HTA-specific details.

The paper is organized as follows.~The data and simulated samples are presented in \cref{sec:data} followed by the Belle II detector description in \cref{sec:detector}.
The initial event selection and reconstruction of the decays are described in \cref{sec:eventsel}.
Corrections introduced to the simulated samples are discussed in \cref{sec:corrs}.
Section \ref{sec:bdt}
details background suppression and final event-selection using machine learning methods. Section \ref{sec:sr} defines the signal region used to extract the \BKnn decay branching fraction. The following two sections, \cref{sec:sigeff} and \cref{sec:background}, are dedicated to the validation of the modeling of the signal-selection efficiency and background contributions, respectively.
Section \ref{sec:stat} documents the statistical approach used to extract the signal and \cref{sec:syst} describes the systematic uncertainties.
The results are discussed in \cref{sec:result},
and consistency checks used for validation are presented in \cref{sec:xcheck}. The combination of the ITA and HTA results is discussed in \cref{sec:combination}. A discussion of the results is presented in \cref{sec:discussion}.
Section \ref{sec:summary} concludes the paper.
\section{Data and Simulated Samples}\label{sec:data}

This search uses data from $e^+e^-$ collisions produced between the years 2019 and 2022 by the SuperKEKB collider \cite{Akai:2018mbz}.~The on-resonance data, with an integrated luminosity of 362\invfb \cite{Abudinen:2019osb}, are recorded at a center-of-mass (c.m.)\ energy of \mbox{$\sqrt{s}=10.58\gev$}, which corresponds to the mass of \Y4S resonance, and contain $N_{\BBbar}=(387 \pm 6) \times 10^6$ $\BBbar$ pairs \cite{Aubert:2004pwa}. 
An additional 42\invfb off-resonance sample, collected at an energy 60\mev below the mass of \Y4S resonance, is used to study background from continuum: $e^+e^- \to \tau^{+} \tau^{-}$ events and  
$e^+e^-\to q\bar{q}$ events, where $q$ indicates an $u$, $d$, $s$, or $c$ quark.

Simulated samples are exploited for training multivariate classifiers, estimating signal-selection efficiencies, identifying backgrounds, and defining components of the fits to data.
Various event generators are used.
The production and decays of charged and neutral \B mesons use \texttt{PYTHIA8}~\cite{Sjostrand:2014zea} and \texttt{EVTGEN}~\cite{Lange:2001uf}.
The \texttt{KKMC} generator~\cite{Jadach:1999vf} is used to generate the $q\bar{q}$ 
pairs followed by \texttt{PYTHIA8} to simulate their hadronization 
and \texttt{EVTGEN} to model the decays of the resulting hadrons.
Similarly, \texttt{KKMC} and \texttt{TAUOLA} \cite{Jadach:1990mz} are employed to simulate production of $e^+e^- \to \tau^{+} \tau^{-}$ events and decays of $\tau$ leptons, respectively. Final-state QED radiation is simulated using \texttt{PHOTOS}~\cite{Barberio:1993qi}.
For all samples, 
the Belle II analysis software 
\cite{Kuhr:2018lps,basf2-zenodo}, interfaced with \texttt{GEANT4} \cite{Agostinelli:2002hh}, is used to simulate the detector response and perform event reconstruction.

The simulated \BKnn signal decays are weighted according to the SM form factor calculations from Ref.~\cite{Parrott:2022zte}.
Similar weighting is applied to $B\to K^*(892) \, \nu\bar{\nu}$ background decays [in the following, $K^*(892)$ mesons are indicated with $K^*$].
The long-distance $B^+\to \tau^+(\to K^+\bar{\nu})\nu$ decays are simulated separately, normalized using the branching fraction from Ref.~\cite{Parrott:2022zte}, and added to the $B^+B^-$ background.
The simulation of several other background processes receives additional corrections.
Nonresonant three-body $B^+\to \Kp n\bar{n} $ decays are simulated assuming the threshold-enhancement effect present in the isospin-partner decay $B^{0}\rightarrow \KS p\bar{p}$~\cite{BaBar:2007esv}. Three-body $B^+\to K^+\KL \KL$ decays are modeled using Dalitz spectra of $B^+ \to K^+ \KS\KS$ decays measured in Ref.~\cite{BaBar:2012iuj} and assuming equal probabilities for the $B^+ \to K^+ \KL\KL$ and $B^+ \to K^+ \KS\KS$ decays. The decay $B^+\to K^+\KS \KL$ is modeled as a sum of a $B^+ \to K^+ \phi(\to \KS\KL)$ resonant contribution and nonresonant p-wave contribution with parameters taken from the isospin-related decay $B^0\to \KS K^+K^-$, as measured in Ref.~\cite{BaBar:2012iuj}. The \texttt{PHOKHARA} event generator~\cite{Czyz:2013xga} is used to simulate $e^+e^-\to\phi(\to \KS\KL)\gamma$ events,
which are used for additional studies.

The simulated continuum samples are normalized based on the known cross sections and integrated luminosity. 
Both the simulated $\BBbar$ background and signal samples are scaled using $N_{\BBbar}$, where the number of $B^+B^-$ pairs 
is calculated as $f^{+-} N_{\BBbar}$, and the number of $\Bz\Bzb$ pairs 
is calculated as $f^{00} N_{\BBbar}$, with $f^{00} = 1 - f^{+-}$ and $f^{+-} = 0.516 \pm 0.012$~\cite{Belle:2022hka}.

\section{Detector}\label{sec:detector}
A comprehensive description of the Belle~II detector is given in Ref.~\cite{Abe:2010gxa}.
The detector consists of several subdetectors arranged in a cylindrical structure around the beam pipe. 
The innermost subsystem consists of a silicon pixel detector surrounded by a double-sided silicon strip detector, referred to as the silicon vertex detector, and a central drift chamber (CDC).  The second layer of the pixel detector covers only one-sixth of the azimuthal angle for the data used in this work.
The silicon detectors allow for precise determination of particle-decay vertices while the CDC
determines
charged-particle momenta and electric charge.
A time-of-propagation counter and an aerogel ring-imaging Cherenkov counter cover the barrel and forward endcap regions of the detector, respectively: these subdetectors are important for charged-particle identification (PID).
An electromagnetic calorimeter (ECL), used to reconstruct photons and distinguish electrons from other charged particles, occupies the remaining volume inside a superconducting solenoid. This provides a uniform  $1.5\,\rm{T}$ magnetic field, 
parallel to the detector's principal axis.
A dedicated system to identify \KL mesons and muons is installed in the
flux return of the solenoid.
The $z$ axis of the laboratory frame is collinear with the symmetry axis of the solenoid and almost aligned with the electron-beam direction.
The polar angle, as well as the longitudinal and transverse directions, are defined with respect to the $z$ axis.

\section{Event Selection} \label{sec:eventsel}
The online-event-selection systems (triggers) for this analysis are based either on the number of charged-particle trajectories (tracks) in the CDC or on the energy deposits in the ECL, and have an efficiency close to $100\%$ for signal decays.
In the offline analysis, the reconstruction of charged particles follows the algorithm outlined in Ref.~\cite{BERTACCHI2021107610}.
For the ITA, to ensure that efficiency is high and well-measured, and to suppress beam-related background, 
charged particles are required to have a transverse momentum $\pt>0.1\gevc$ and to be within the CDC acceptance ($17^\circ<\theta<150^\circ$). 
All charged particles except those used to form \KS candidates are required 
to have minimum longitudinal and transverse distances (impact parameters) from the average interaction point
of $|\dz|<3.0\cm$ and $\dr<0.5\cm$, respectively. 
The \KS candidates are formed by combining pairs of oppositely charged particles in a vertex fit. These candidates are required to have a dipion reconstructed mass between $0.495$ and $0.500$ \gevcc, vertex $p$-value greater than 0.001, flight time greater than 0.007~ns (corresponding to about 2~mm displacement from the primary vertex), and cosine of the angle between momentum and flight direction greater than 0.98.
Photons are identified as energy deposits exceeding $0.1\gev$ detected in the ECL regions within the CDC acceptance, and not matched to tracks. The minimum energy requirement suppresses the beam-related background and energy deposits from charged hadrons that fail the matching to tracks.
Each of the charged particles and photons is required to have an energy of less than 5.5\gev to reject misreconstructed particles and cosmic muons. The kaon candidates are selected using particle-identification likelihoods based on 
information coming primarily from the PID detectors, complemented with information from the silicon strip detector, CDC, and the $\KL$ and muon identification system.~To ensure reliable PID, at least 20 deposited-charge measurements are required in the CDC. 
The chosen PID requirement has 68\% efficiency for signal kaons, while the probability to identify a pion as a kaon is 1.2\%.  Candidates are also required to have at least one deposit in
the pixel detector: this improves the impact parameter resolution, and helps to
reject background events. 

Events are required to contain no more than ten tracks to suppress background (e.g.,~high-multiplicity continuum production) with only a 0.5\% loss of signal-selection efficiency.~Low-track-multiplicity background events, such as those originating from two-photon-collision processes, are suppressed by demanding at least four tracks in the event. This reduces signal-reconstruction efficiency by $7.6\%$.
The total energy from all reconstructed particles in the event must exceed 4\gev. The polar angle of the missing momentum, computed in the c.m.\ frame as the complement to the total momentum of all reconstructed particles, must be between $17^\circ$ and $160^\circ$. 
This range is chosen to remove low-multiplicity events and to ensure that the missing momentum points toward the active detector volume.

To select the signal kaon in an event, the mass squared of the neutrino pair is computed as 
\begin{equation}
q^2_{\mathrm{rec}} = s/(4 c^4) + M^2_K- \sqrt{s} E^*_K / c^4
\end{equation}
assuming the signal \B meson to be at rest in the $e^{+}e^{-}$ c.m.\ frame.
Here $M_K$ is the known mass of $K^+$ mesons and $E^*_K$ is the reconstructed energy of the kaon in the c.m.\ system.
Uncertainties in the kinematic properties of the colliding beams have negligible impact on the $q^2_{\mathrm{rec}}$  reconstruction.
 The candidate having the lowest $q^2_{\mathrm{rec}}$ is retained for further analysis. 
Studies on simulated signal events show that prior to applying the $q^2_{\mathrm{rec}}$ requirement the fraction of events with multiple candidates is 39\%. The average number of candidates in such events is 2.2.
The lowest-$q^2_{\mathrm{rec}}$ candidate is the signal kaon in 96\% of cases.
Checks using a random selection of the signal candidate, if several candidates are found, indicate no bias in the procedure.
The remaining charged particles are fit to a common vertex and are attributed, together with the photons and \KS candidates, to the rest of the event (ROE).
For the signal events, these charged particles and \KS candidates correspond to the decay products of the second \B meson.

The HTA commences with the full reconstruction of a $\B$ meson (\Btag), decaying into one of 
36 hadronic $B$ decays, through the full event interpretation (FEI)~\cite{Keck:2018lcd}.
The FEI is an algorithm based on a hierarchical multivariate approach in which
final-state particles are constructed using the tracks and energy deposits in the ECL, and combined into intermediate particles until the final \Btag candidates are formed.
The algorithm calculates, for each decay chain, the probability of it correctly describing the true process using gradient-boosted decision trees. Only \Btag mesons with a probability exceeding 0.001 are retained.
In addition, the beam-constrained mass $M_{\mathrm{bc}}=\sqrt{s/(4c^4)-{p^*_B}^2/c^2}>5.27\gevcc$ and 
$\left|\Delta E\right|=\left|E^*_B-\sqrt{s}/2\right|<300\mev$ are
required, where $E^*_B$ and $p^*_B$ are the energy and the magnitude of the three-momentum of the \Btag in the c.m.\ frame, respectively. Signal candidates peak at the known $\Bp$ mass and zero in $M_{\mathrm{bc}}$ and $\Delta E$, respectively, while continuum events are distributed more uniformly.
The FEI algorithm imposes conditions on the charged particles and energy deposits in the ECL similar to those used by the ITA. The algorithm requires at least three tracks and three energy 
deposits in the ECL, including those that are associated with the tracks. 
Furthermore, events with more than 12 tracks having $d_{r} < 2$~cm and $|d_{z}| < 2$~cm are rejected. Such events would have greater multiplicity than the maximum of the reconstructed \Btag final states plus the signal-kaon track.

The signal-kaon candidate track is required to have at least 20 measurements in the CDC, and impact parameters $d_{r}<0.5$~cm and $|d_{z}|<4$~cm, and is required to satisfy PID criteria for a kaon. The \Btag and signal kaon are required to have opposite charges.
The same restrictions on missing momentum are applied as in the ITA.
Moreover, the number of tracks with $d_{r}<2$ cm, $|d_{z}|<4$ cm and with at least 20 measurements in the CDC, which are neither associated with the \Btag nor with the signal kaon, is required to be zero.

The remaining reconstructed objects in the HTA include tracks, which neither meet the CDC nor impact parameter requirements (extra tracks), and energy deposits in the ECL, which are neither associated with the \Btag
 nor the signal kaon (extra photon candidates). 
Only the energy deposits in the ECL that exceed a $\theta$-dependent energy threshold ranging from $60$~\mev to $150$~\mev, have a distance from the nearest track extrapolation larger than 50~\cm, and are reconstructed within the CDC acceptance are considered. The sum of the energies of these deposits, denoted as \Eecl, and the multiplicity of the extra tracks, denoted as $n_{\rm{tracks~extra}}$, are utilized in the subsequent steps of the analysis. 
Events are rejected if a $\KS$-meson, $\piz$-meson, or $\Lambda$-baryon candidate is reconstructed from the extra tracks and photons.

\section{Corrections to simulated data} \label{sec:corrs}
The simulation of the detector response is tested using control samples from data, and correction factors are introduced with corresponding systematic uncertainties. 
Correction factors are applied as weights to the selected events when appropriate, particularly when the corrections impact the efficiency of the signal-kaon selection.
In other cases, when the corrections affect the kinematic properties of the particles, these corrections are applied prior to the event selection and computation of 
related variables.
The correction procedure is carried out for the nominal analysis
as well as when computing systematic variations.

\subsection{Reconstruction of charged particles}
The efficiency for reconstructing charged particles is studied using $e^+ e^- \to \tau^+\tau^-$ events, where one $\tau$ lepton decays into a single charged particle while the other decays into three charged particles \cite{Collaboration:2035}. Simulation agrees well with the data. A systematic uncertainty of $0.3\%$ is introduced for each charged particle to account for uncertainties in the detection efficiency and in the knowledge of the detector geometrical acceptance.

The reconstruction of kinematic properties of charged particles is validated by comparing the measured pole masses of known resonances with simulation. 
The simulation reproduces the data with an accuracy better than $0.1\%$, and any residual differences have a negligible impact on the analysis.

\subsection{Identification of charged particles}
About 10\% of background arises from incorrect particle identification of the 
signal-kaon candidate.
The main contribution is from misidentified pions, while misidentified muons, electrons, and protons have a smaller impact. The efficiency of kaon identification and the misidentification (``fake'') rate for pions misidentified as kaons are determined using
$D^{*+} \to \pi^+ \Dz(\to \Km\pi^+)$ decays reconstructed in continuum data and simulation. The small mass difference between $D^{*+}$ and $D^0$ mesons enables the isolation of  a pure signal. The charge of the low-momentum pion from the flavor-conserving $D^{*+}$ decay allows the precise identification of the products of the Cabibbo-favored \Dz decay, providing abundant and low-background $K^-$ and $\pi^+$ samples.

Correction factors and their uncertainties are applied to the simulation as functions of the particle's charge, momentum, and polar angle.
The correction factors for the pion-to-kaon fake rates are close to a factor of 2, indicating that the simulation underestimates the rate at which pions are misidentified as kaons. The uncertainties associated with these corrections, which are around $1\%$ for efficiencies and $10\%$ for fake rates, are treated as systematic uncertainties. Correction factors for the lepton-to-kaon fake rates are also applied, 
although their impact is negligible.

The correction factors for kaon identification efficiency and the pion-to-kaon fake rate are further validated for the signal region of the ITA, using the $\Bp \to h^+\Dzb(\to K^+\pi^-)$ decays, where $h^+$ stands for $\pi^+$ or $\Kp$, following the procedure outlined below. 
\begin{figure}
    \centering
    \includegraphics[width=\linewidth]{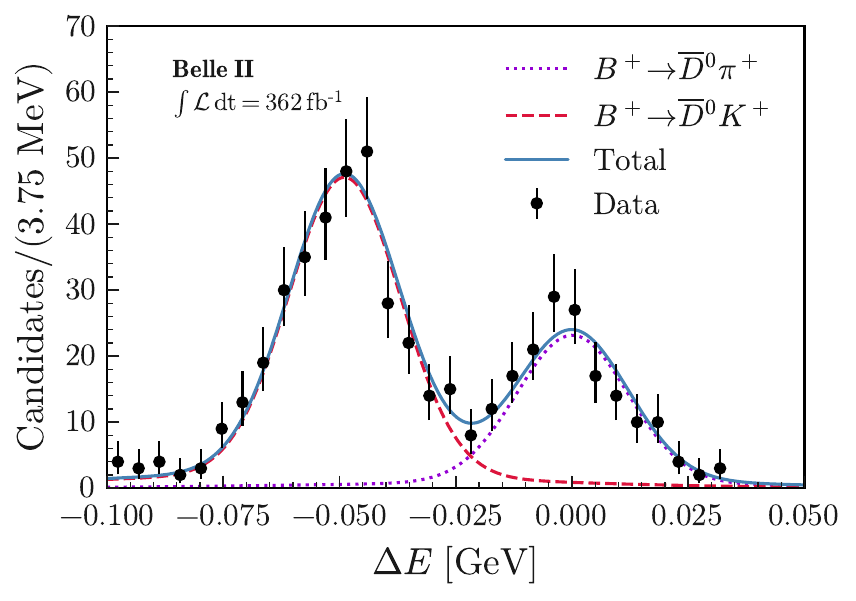}
    \caption{Distribution of $\Delta E$ in data (dots with error bars) obtained for $B^+ \to h^+\Dzb$ decays, where $h^+ = \pi^+$ or $K^+$,
    computed assuming a pion mass hypothesis for $h^+$. 
    The blue solid line represents the fit result to the data, modeled as a sum of two Gaussian shapes corresponding to $B^+\to K^+\Dzb$ (dashed red line) and $B^+\to \pi^+\Dzb $ (dotted magenta line) decays.
     Events selected for the figure are
     reconstructed as $\Bp \to K^+ \nu\bar{\nu}$ events, with the daughters from the $\Dzb$ decays removed, and chosen to be in the signal region of the ITA.}
    \label{fig:pidtestm}
\end{figure}
The decays are reconstructed using the pion mass hypothesis and the nominal kaon identification
for the $h^+$ candidate.
For $\Bp \to \pi^+\Dzb$ decays, the distribution of the $\Delta E$ variable peaks at zero.
Since the $\Bp$ is produced almost at rest in the c.m. frame, the kaon momentum in the two-body $\Bp \to K^+\Dzb$ decay is expected to be equal to $2.3\gevc$ with a small spread. Combined with the pion mass, this leads to an energy deficit compared to the correct kaon hypothesis, resulting in a shift in $\Delta E$ of $-0.049 \gev$.
This characteristic of the $\Delta E$ distribution is used to distinguish between $\Bp \to K^+\Dzb$ and $\Bp \to \pi^+\Dzb$ decays without relying on PID information.
The kaon and pion candidates from the $\Dzb$ decay are differentiated by kaon identification: the particle with the higher value is assumed to be a kaon. 
Only $\Dzb$ candidates with an invariant mass within 3 standard deviations of the known \Dzb mass~\cite{ParticleDataGroup:2022pth} are kept. In addition, the selections $M_{\rm{bc}}>5.27 \gevcc$ and $|\Delta E|< 0.1 \gev$ are applied. If several candidates pass the selection, a random one is chosen. 

For the selected $\Bp \to h^+\Dzb$ decays, the information on the $\Delta E$ variable is kept while the tracks from the $\Dzb$ decay are removed and each event is reconstructed again as a $\Bp \to K^+ \nu\bar{\nu}$ event. The same procedure is repeated for both data and simulation. The selected events show a $q^2_{\mathrm{rec}}$ distribution peaking between $3\gevgevcccc$ and $5\gevgevcccc$ corresponding to the \Dz mass squared. The events have a $\Bp \to \Kp \nu\bar{\nu}$ signal-like signature and for this $q^2$ range are reconstructed with high efficiency. Distribution for $q^2$ is included in the Supplemental Material~\cite{supplemental}.
The distribution of the $\Delta E$ variable for the signal region of the ITA is shown in \autoref{fig:pidtestm}. Two prominent peaks corresponding to $h^+ = \Kp$ and $h^+ = \pi^+$ are observed. The yields of the two components are extracted in a fit
using Gaussian shapes.  
The double ratio of the $\Bp \to \pi^+ \Dzb$ to $\Bp \to K^+\Dzb$ decay rates in data to simulation is $1.03\pm 0.09$, showing consistency with unity within the statistical uncertainty.

\subsection{Reconstruction of neutral particles} \label{sec:neutral}
The photon detection efficiency at Belle II, and the calibration of the
photon energy reconstruction, are based on studies of $e^+ e^- \to \mu^+
\mu^- \gamma$ events. The efficiency is the fraction of events where a
photon is reconstructed with momentum consistent with the expectation
from recoil against the $\mu^+ \mu^-$ system~\cite{Henrikas:3493}. The resulting
uncertainty is negligible for this analysis.

The uncertainty on the photon energy is 0.5\%. The effect on signal
yield in the fit is estimated by applying this uncertainty
to energy deposits from photon candidates matched to simulated photons.

The simulated sample shows that photon candidates have $30\%$ contamination
from beam-related background, energy deposits from charged hadrons that are reconstructed away from the particle trajectory, and from neutral hadrons. These deposits are not matched to simulated photons (\lq\lq unmatched\rq\rq). 
The bias in the reconstructed energy for these sources 
(\lq\lq hadronic energy correction\rq\rq) is studied using the summed energy of the photon candidates in the ROE  (``summed neutral energy'', $\sum E_{\gamma}$) of events containing a $B^+\to K^+ \jpsi$ decay (details on $B^+\to K^+ \jpsi$ decay reconstruction can be found in \cref{sec:sigeff}). 
In the simulation, the energy of reconstructed photon candidates is treated differently based on their matching to the generated photons.
The energy for matched candidates is not corrected.
For unmatched candidates, a multiplicative hadronic energy correction
is inferred empirically using data.
In the simulation the 
correction is varied within a $\pm 20\%$ range around unity. 
For the ITA, an improvement is found when the hadronic energy is varied down by $10\%$. The corresponding correction with $100\%$ uncertainty (relative) is introduced. Illustration of the hadronic energy correction for the ITA is included in the Supplemental Material~\cite{supplemental}.

Figure \ref{fig:energy_scale} shows the comparison of distributions of summed neutral energy for events in which  a $B^+\to K^+ \jpsi$ decay is reconstructed, for collision data and for the corresponding uncorrected and corrected simulation. The correction corresponds to a variation of the hadronic energy by $-10\%$. Better data-simulation agreement is achieved by the corrected simulation.

The correction is validated using various control samples dominated by background, such as off-resonance data and data at early selection steps. An improvement is observed in the description of several variables related to neutral-particle energy deposits, such as the number of photon candidates. The latter is sensitive to the hadronic energy since the hadronic-energy deposits peak at low energy and are affected strongly by the minimal energy requirement of $0.1\gev$.

For the HTA, a different extra-photon selection is adopted. 
In the HTA sample, the energy spectrum of extra photon candidates exhibits good data-simulation agreement, 
but observed discrepancies in the multiplicity (\nge) propagate to the \Eecl distribution.
To correct this, a control sample is used where the signal kaon and the \Btag have the same charge. A weight is computed for the $\nge$ distribution as follows:
\begin{equation}
w_{\nge} = \frac{N_\textrm{data}(\nge)}{N_\textrm{simulation}(\nge)},
\end{equation}
where $N_{\text{data}}(\nge)$ and $N_{\text{simulation}}(\nge)$ correspond to the event yields with \nge candidates in data and simulation, respectively. Subsequently, simulated events where the signal kaon and \Btag have opposite charges, are weighted based on their associated \nge value.

\begin{figure*}[htp]
\centering
\includegraphics[width=0.49\linewidth]{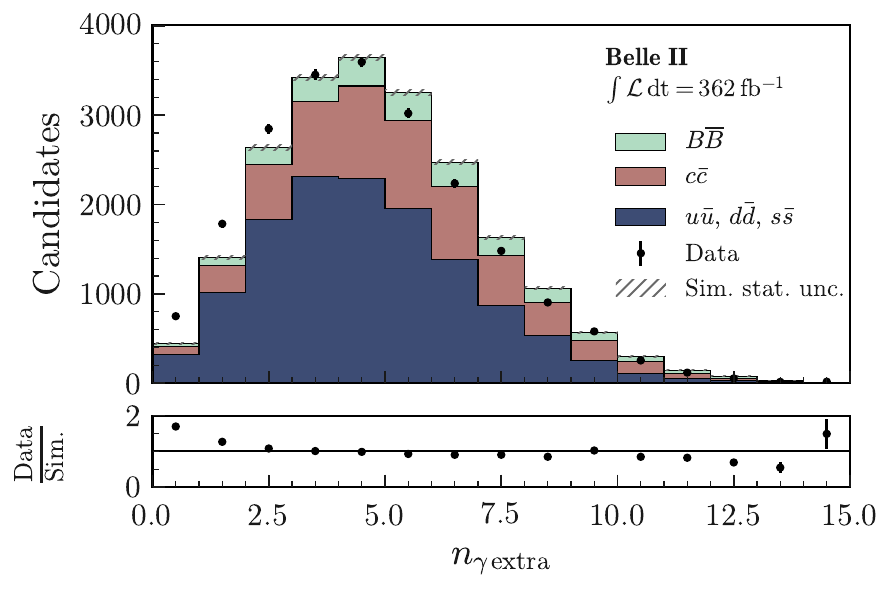}
\includegraphics[width=0.49\linewidth]{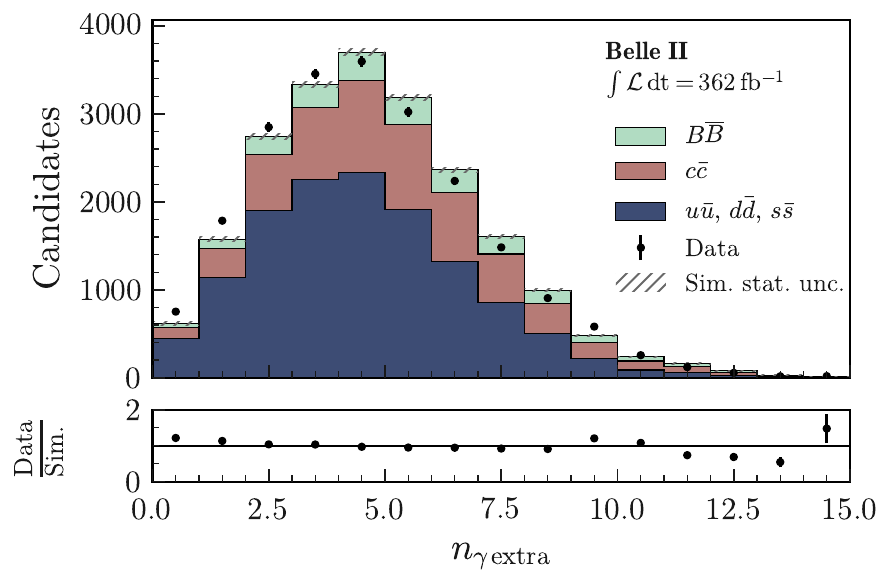}
\caption{Distributions of  the number of extra photon candidates in the HTA after the selection described in~\cref{sec:eventsel} in data (points with error bars) and simulation (filled histograms) for the opposite-charge pion-enriched control sample, on the left before the photon multiplicity correction and on the right after the correction. The yields are shown
individually for the three background categories ($\BBbar$ decays, $c\bar{c}$ continuum, and light-quark
continuum). The data-to-simulation ratios are shown in the bottom panels. }
\label{fig:ngamma_HTA}
\end{figure*}

This method is validated using an independent pion-enriched control sample where the signal track is identified as a pion instead of
a kaon.
The pion-enriched sample is further divided into two samples based on whether the signal candidate and \Btag have the same or opposite charge. Corrections are derived from the sample where signal and \Btag have same charge
and are then applied to the opposite-charge sample. The effect of the correction in the pion-enriched sample at the event-selection stage is shown in \cref{fig:ngamma_HTA}.
\begin{figure}[htp]
    \centering
    \includegraphics[width=\linewidth]{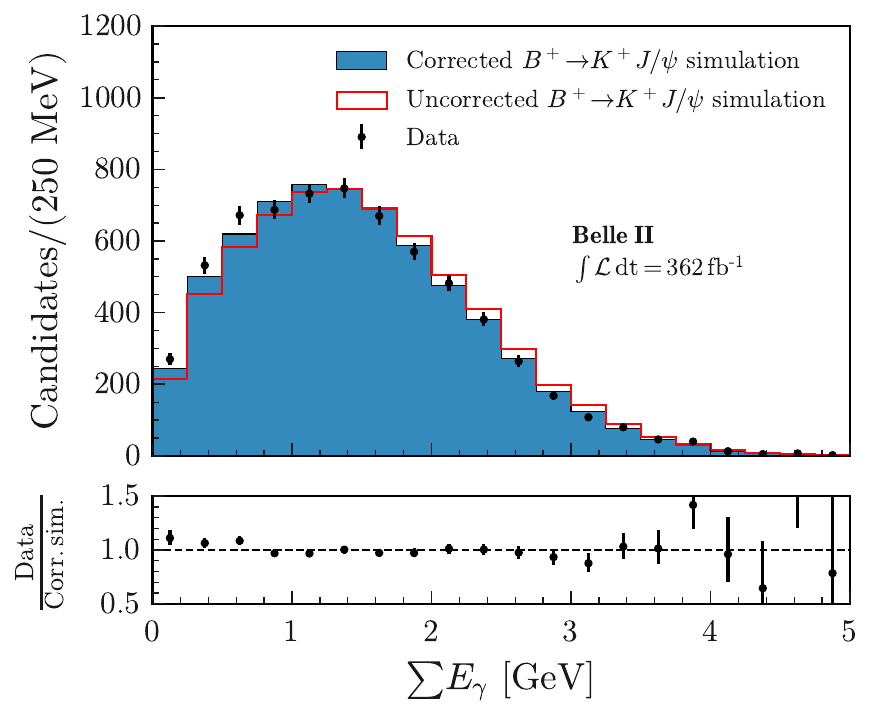}
    \caption{Distribution of the summed energy of the photon candidates obtained in collision data (points with error bars), uncorrected simulated data (open histogram), and corrected simulated data (filled histogram), for events in which a $B^+\to K^+ \jpsi$ decay is reconstructed.
    The correction corresponds to a variation of the hadronic energy by $-10\%$. The simulation is normalized to the number of events in data. The ratio shown in the lower panel refers to data over corrected simulation.}
    \label{fig:energy_scale}
\end{figure}

Although an improvement is observed after applying the correction, residual data-simulation discrepancies remain. To account for these, a systematic uncertainty is assigned corresponding to $100\%$ of the residual difference in the data-to-simulation ratio observed in the opposite-charge pion-enriched  control sample after the correction.

\label{sec:klmod}
\begin{figure}[htp]
    \centering
    \includegraphics[width=\linewidth]{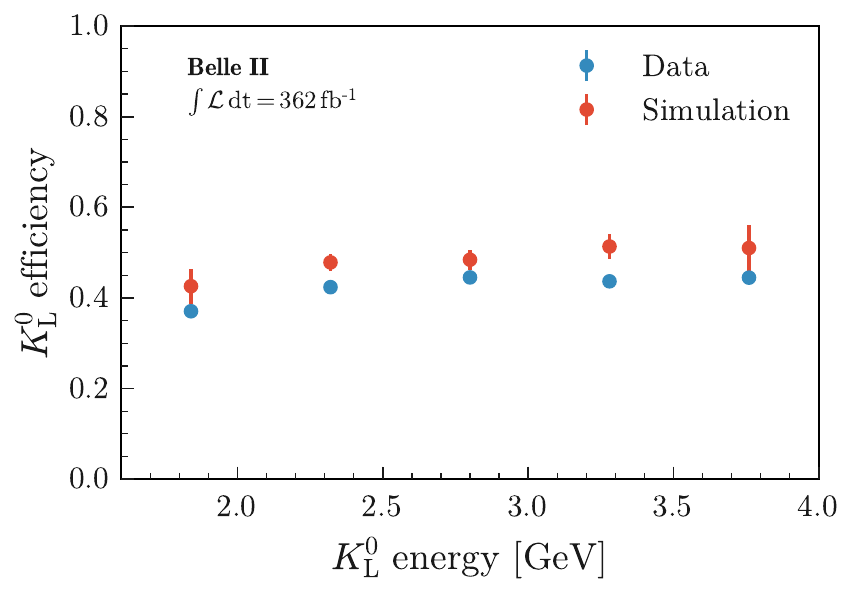}
    \caption{Efficiency of reconstructing an energy deposit in the ECL matched to the \KL direction, as a function of the \KL energy, for $e^+e^- \to \gamma \phi$ data and simulation. The energy deposits are selected following the ITA criteria.}
    \label{fig:kl}
\end{figure}
Given the prominence of background contributions containing \KL mesons, a dedicated study is performed to check their modeling.
This study focuses on the ECL response only, as the analysis does not use $\KL$ candidates from the dedicated identification system to avoid additional systematic uncertainties due to their modeling.
Radiative-return production $e^+e^- \to \gamma \phi (\to \KS\KL)$ is used for this purpose for \KL with energy above 1.6 \gev.
 The events are selected by demanding a photon candidate with energy $E^*_{\gamma}>4.7$~\gev in the c.m.\ frame, a well-reconstructed \KS candidate, and no extra tracks. The \KL four-momentum is inferred based on the photon and \KS four-momenta, where the photon energy is computed based on the two-body $e^+e^- \to \gamma \phi$ process. The typical momentum resolution of an inferred \KL is better than 1\%.  
An energy deposit in the ECL reconstructed at a radius $R$ is matched to the trajectory of \KL extrapolated to the same 
$R$ if the distance between them is less than 15\cm.
The efficiency for finding a matched energy-deposit is studied both in data and simulation,
 and is tested separately in the ITA and HTA.
 The ITA selection for the ECL deposits is looser than the HTA selection; therefore a higher efficiency is found. 
 Figure~\ref{fig:kl} shows the ITA \KL efficiency as a function of momentum; the simulation overestimates the efficiency by 17\%.
This is taken into account by performing a $-17\%$ (relative) efficiency correction in the ITA sample, for all \KL, including those below 1.6 \gev.
A $\pm 8.5\%$ systematic uncertainty (i.e.\ half of the correction) is assigned.
Distribution of the energy deposits in ECL is shown in the Supplemental Material~\cite{supplemental}.

While the radiative-return production of $\phi$ mesons does not encompass $\KL$ with energies below $1.6~\gev$, approximately half of the \KL mesons in the main background processes populate this lower-energy range.
As a consistency check, a $100\%$ inefficiency is incorporated in the ITA for this kinematic region in the simulation.
Specifically, all energy deposits in the ECL that fall within a $15$~cm radius of the extrapolated $\KL$ trajectory are removed for simulated $\KL$ with energies smaller than $1.6~\gev$. The impact of this additional requirement on the analysis is found to be covered by the hadronic-energy systematic uncertainty, discussed above. 

The \KL reconstruction efficiency is smaller for the HTA. Since the effect on \Eecl is already addressed by the correction and systematic uncertainty derived from the extra-photon-multiplicity spectrum, no direct correction to the \KL efficiency is applied. Instead, a systematic uncertainty is assigned, wherein the yields of \B final states with a \KL are varied by 17\%.

\section{Background Suppression} \label{sec:bdt}
Simulated signal and background events are used to train BDTs that suppress the background.
Several inputs are considered,
including general event-shape variables described in Ref.~\cite{Bevan:2014iga}, as well as variables characterizing the kaon candidate and the kinematic properties of the ROE.
Moreover, vertices of two and three charged particles, 
with one of the tracks being the kaon-candidate track, 
are reconstructed to identify kaons from \Dz and \Dp meson decays; variables describing the fit quality and kinematic properties of the resulting candidates are considered as possible BDT inputs.
Variables are excluded if either their contribution to the classification's 
separation power is negligible or they are poorly described by the simulation.

The ITA uses two consecutive BDTs. A first binary classifier, \BDT1, is designed as a first-level filter after event selection. It is trained on $10^6$ simulated events  
of each of the seven considered background categories
(decays of charged \B mesons, decays of neutral \B mesons, and the five continuum categories: $e^+e^- \to q\bar{q}$ with $q=u,d,s,c$ quarks and $e^+e^- \to \tau^{+} \tau^{-}$), 
weighted to a common equivalent luminosity such that the sum of weights is balanced to the $10^6$ simulated signal events.
The classifier uses 12 input variables. The most discriminating variable is the difference between the ROE energy in the c.m.\ frame and $\sqrt{s}/2$
($\Delta E_\textrm{ROE}$), which tends to be negative for signal events due to neutrinos, whereas it is positive for the background with additional reconstructed particles. 
Significant discrimination comes from variables sensitive to the momentum imbalance of the signal events due to neutrinos, as well as those that correlate the missing momentum with the signal-kaon momentum. Examples of such variables are the reduced first-order Fox-Wolfram moment \cite{Fox:1978vw} and the modified Fox-Wolfram moments \cite{PhysRevLett.91.261801}.

The second classifier, \BDT2, is used for the final event selection.
It is trained on events with \BDT1$>0.9$, which corresponds to a signal (background) selection efficiency of 34\% (1.5\%),
using 35 input variables.
A simulated background sample of
200\invfb equivalent luminosity, corresponding to $4.2\times 10^6$ events, and a sample of $1.7\times 10^6$ signal events are used.
Tests with larger samples used for \BDT2 training show no additional improvements in \BDT2 performance.
For \BDT2, the most discriminating variables are
the cosine of the angle between the momentum of the signal-kaon candidate and the thrust axis of the ROE computed in the c.m.\ frame, which has a uniform distribution for the signal and a peaking shape for the jetlike continuum background. The thrust axis is defined as the unit vector $\hat {t}$ that maximizes the thrust value $\sum |\hat{t} \cdot \vec{p}_{i}^{\ *}|/\sum |\vec{p}_{i}^{\ *}|$, where $\vec{p}_{i}^{\ *}$ is the momentum of $i$th final-state particle in the $e^{+}e^{-}$ c.m.\ frame~\cite{Brandt:1964sa,Farhi:1977sg}.
Also important are variables identifying kaons from \Dz and \Dp meson decays, and the modified Fox-Wolfram moments.
The \BDT1 and \BDT2 parameters are optimized based on a grid search in the parameter space and are described in \cref{app:ita_vars}. Training of \BDT1 and \BDT2 classifiers is based on simulated samples that are statistically independent of those used   
in the sample-composition fit. 

For the HTA, the remaining background is suppressed using a multivariate classifier BDTh, which uses 12 input variables combining information about the event shape, the signal-kaon candidate, the \Btag meson, and any extra tracks and extra photons.
Simulated background samples of about $2 \times 10^5$ $\BBbar$ events and $3 \times 10^5$ continuum events, which correspond to an equivalent luminosity of, respectively, 3~$\invab$ and 1~$\invab$, are used together with a signal sample of $5 \times 10^5$ events.
The BDTh parameters are optimized through a grid search in the parameter space. 
Given the limited size of the simulated sample, it is beneficial to use information from the whole sample both to train the BDTh and estimate the remaining background in the signal region. The simulated sample is thus split into two subsamples that are used to train two separate BDTh's.
Good agreement between the two outputs is observed. 
The data sample is then randomly divided into two halves and each BDTh is applied to one half.
In the background sample, for each event, the BDTh other than the one the event is used to train is applied. 
Details regarding the input variables and BDTh parameters are reported in \cref{app:hta_vars}.

The BDTh input variable providing the highest discriminating power is \Eecl.
For correctly reconstructed signal events, no extra ECL deposits are expected, which results in a \Eecl distribution peaking at zero; backgrounds leave deposits with energies up to 1~\gev. The second most discriminating variable is the sum of missing energy and magnitude of the missing momentum (\sumEp), where the missing four-vector is defined as the difference between the beam four-vector
and the sum of the signal kaon and \Btag four-vectors in the c.m.\ frame. For correctly reconstructed signal events, \sumEp is defined by the neutrino kinematic properties, and its distribution peaks around 5~\gev, while for background events the random loss of particles mimicking the neutrinos results in a broader distribution.

\section{Signal Region Definition} \label{sec:sr}
Using the simulated signal sample, the \BDT2 variable is mapped to the complement of the integrated signal-selection efficiency,
\begin{equation}
  \eta(\BDT2) = 1 - \int_{\BDT2}^{1} \xi(b) \mathrm{d} b\,,
\end{equation}
where $\xi(b)$ is the total signal-selection efficiency density for the \BDT2 value $b$. In this way the distribution of $\eta(\BDT2)$ for simulated signal events is uniform;
a similar mapping is used to define $\eta(\mathrm{BDTh})$, based on
the efficiency of the selection on BDTh.

For the ITA, the signal region (SR) is defined to be $\BDT1>0.9$ and $\eta(\BDT2)>0.92$, as this criterion maximizes the expected signal significance, based on studies in simulation. The SR is further divided into $4\times3$ intervals (bins) in the $\eta(\BDT2)\times q^2_{\mathrm{rec}}$ space. 
The bin boundaries are $[0.92, 0.94, 0.96, 0.98, 1.00]$ in $\eta(\BDT2)$ and $[-1.0, 4.0, 8.0, 25.0]\gevgevcccc$ in $q^2_{\mathrm{rec}}$. 
The bin $\eta(\BDT2)>0.98$ provides the main information on the signal while the bin $\eta(\BDT2)<0.94$ helps to constrain background contributions. 
The bin boundaries in $q^2_{\mathrm{rec}}$ are chosen to follow those of theoretical predictions \cite{Buras:2014fpa} while ensuring a sufficient number of expected signal events in each bin.
The expected yields of the SM signal and the backgrounds in the SR are 160 and 16793 events, respectively. More detailed information about the expected background composition for charged and neutral $B$ decays is shown in the Supplemental Material~\cite{supplemental}.
For the highest-purity $\eta(\BDT2)>0.98$ region, the expected SM signal yield is reduced to 40 events with a background yield of 977 events. These signal and background yields include corrections to the simulation discussed in the following sections; they correspond to the sample entering the statistical analysis to extract the signal described in \cref{sec:stat}.

For the HTA, the SR is defined to be $\eta(\mathrm{BDTh})>0.4$ and is divided into six bins with bin boundaries at $[0.4, 0.5, 0.6, 0.7, 0.8, 0.9, 1.0]$. In events containing multiple \Btag-$K^{+}$ candidates, the candidate formed by the \Btag with highest FEI probability is selected.
The expected yields of the SM
signal and the background in the SR are 8 and 211 events, respectively.
For the highest purity $\eta(\mathrm{BDTh})>0.7$ region, the expected SM signal yield is reduced to 4 events
with background yield of 33 events.

The expected background and signal distributions in the signal search region are shown in the Supplemental Material~\cite{supplemental}.

\begin{figure*}[ht]
        \centering
\includegraphics[width=0.49\linewidth]{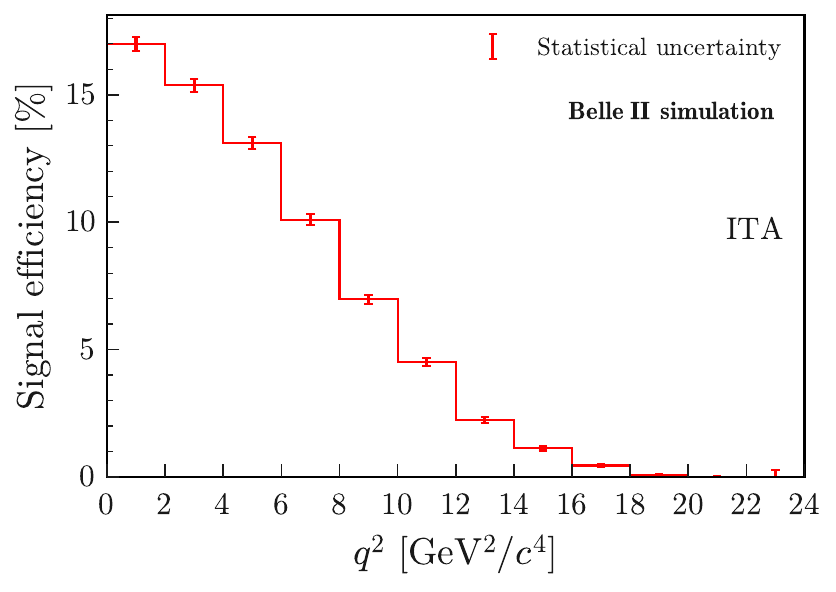}
\includegraphics[width=0.49\linewidth]{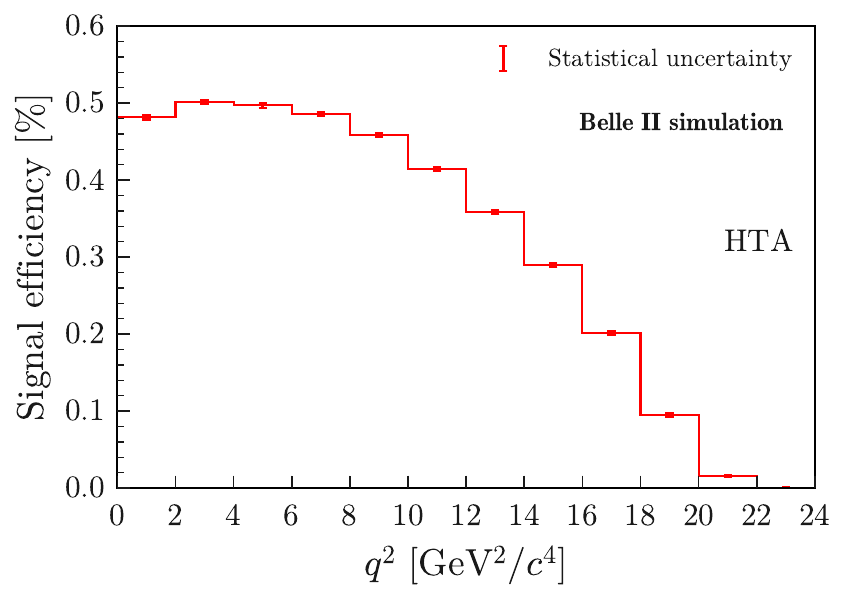}
    \caption{Signal-selection efficiency as a function of the dineutrino invariant mass squared $q^2$ for simulated events in the SR for the ITA (left) and HTA (right). The error bars indicate the statistical uncertainty.}
    \label{fig:sigeff}
\end{figure*}

The signal-selection efficiency in the SR is shown in~\cref{fig:sigeff}. Much higher efficiency is observed for the ITA; however, the ITA efficiency has a significantly stronger $q^2$ dependence compared to the efficiency for the HTA. The analysis relies on modeling of this variation by simulation, which is checked using a control channel, as discussed in the next section.

\section{Signal selection efficiency validation} \label{sec:sigeff}
The decay  $\BJpsiK(\tomumu)$ is used
to validate the BDT performance on signal-like events between data and simulation, exploiting its large branching fraction and distinctive experimental signature.
These events are selected in data and \BJpsiK simulation by requiring the presence of two oppositely-charged muons with an invariant mass within 50\mevcc of the known \jpsi mass \cite{ParticleDataGroup:2022pth}.~To suppress background events, the variable $\left|\Delta E\right|$ is required to be less than 100\mev and the beam-energy constrained mass $M_{\mathrm{bc}}$ is required to exceed 5.27\gevcc.
These criteria result in $7214$ events being selected in the data sample with an expected background contamination of $2\%$.
Each event is then reconsidered as a \BKnn event by ignoring the muons from the \jpsi decay and replacing the kaon candidate with the
signal kaon candidate from a simulated \BKnn event, to reflect the three-body topology of the signal signature.
The kinematic properties of the signal kaon are then adjusted such
that the $B^+$ four-momentum and decay vertex in the simulated \BKnn decay match the four-momentum and decay vertex of the corresponding $B^+$ from the \BJpsiK decay.
This substitution is performed for the reconstructed track, ECL energy deposits, and PID likelihood values 
associated with the simulated kaon such that the test samples have a format identical to the data and can be analyzed by the same reconstruction software. 
This signal-embedding method is performed for both data and \BJpsiK simulation.

\begin{figure}[htp]
\centering
\includegraphics[width=\linewidth]{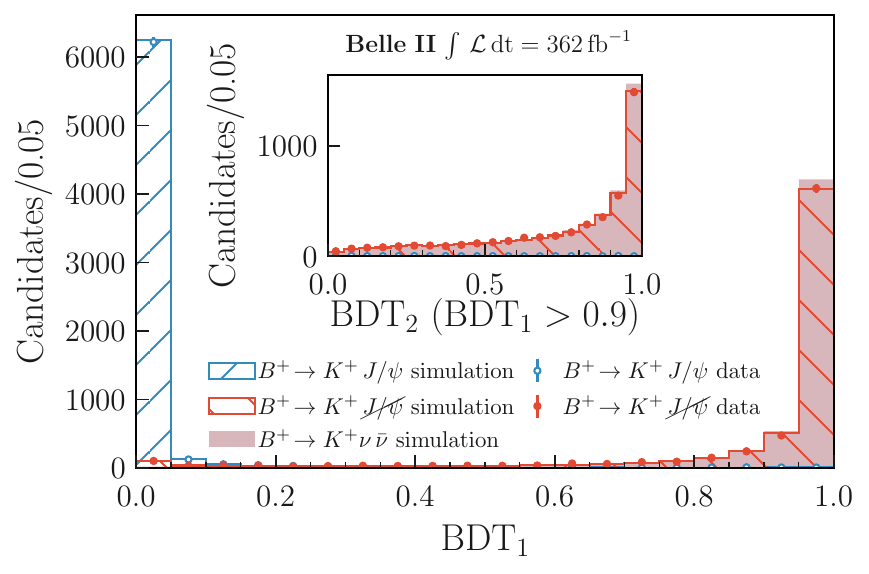}
\caption{
	Distribution of the classifier output \BDT1 (main figure) and \BDT2 for $\BDT1>0.9$ (inset).
	The distributions are shown before (\BJpsiK) and after (\BJpsiKSlash) the muon removal and replacement of the kaon momentum of selected \BJpsiK events in simulation and data.
	As a reference, the classifier outputs directly obtained from simulated \BKnn signal events are overlaid.
	The simulation histograms are scaled to the total number of \BJpsiK events selected in the data.
	}
\label{fig:jpsi_bdt}
\end{figure}

The results obtained by analyzing selected events are summarized for the ITA in \cref{fig:jpsi_bdt}, where the distributions of the output values of both BDTs are shown.
Good agreement between simulation and data is observed for the selected events before 
(\BJpsiK) and after (\BJpsiKSlash) the signal embedding. Distributions with logarithmic $y$-axis are presented in the Supplemental Material~\cite{supplemental}.
The ratio of the selection efficiencies for the SR in data and simulation is  $1.00\pm0.03$; i.e., agreement is observed.

For the HTA, the signal embedding is used to check both the FEI and the combined FEI plus BDTh signal reconstruction efficiency. The ratios of data and simulation efficiencies at the two levels of the selection are found to be $0.68\pm0.06$ and $0.60 \pm0.10$, respectively.
The first ratio agrees with an independent FEI calibration derived from $\B\to X \ell\nu$ FEI-tagged events~\cite{Belle-II:2020fst} and is therefore used as a correction for signal efficiencies and $\BBbar$
normalization.
From the relative uncertainty on the efficiency ratio computed after the $\eta(\textrm{BDTh})$ selection, a 16\% systematic uncertainty on the signal-selection efficiency is derived.
For HTA, the resulting distributions of this study are shown in the Supplemental Material~\cite{supplemental}.

\section{Background studies} \label{sec:background}
The main background sources for the analysis arise from decays that involve an energetic kaon (or a misidentified pion), missing energy, or particles that leave no or small signatures in the ECL, such as \KL mesons. These processes occur in both continuum and $B$-meson decays.
Dedicated studies, using a variety of control samples, are performed in order to validate the background description in simulated events. Where needed, correction factors are derived with corresponding systematic uncertainties. In the following subsections the modeling of backgrounds from continuum (\cref{sec:continuum}) and $\BBbar$ events (\cref{sec:BBbar}) are discussed. In \cref{sec:normB} the overall background normalization after all corrections are applied is checked.

\subsection{Continuum background} \label{sec:continuum}
Continuum represents 40\% and 30\% of the background in the entire signal region of the ITA and HTA, respectively. This contribution drops to $17\%$ in the highest-sensitivity region $\eta(\BDT2)>0.98$ of the ITA, and to $15\%$ in the highest-sensitivity region $\eta(\mathrm{BDTh})>0.7$ of the HTA. The background modeling is validated using the off-resonance data and shows moderate disagreements in the shape of some of the input features of the various classifiers (locally up to 20\%).
The modeling of continuum-background simulation is thus improved following Ref.~\cite{Martschei_2012}.
A binary classifier, BDT$_\mathrm{c}$, is trained to separate the off-resonance data and off-resonance simulation.
For the ITA, the BDT$_\mathrm{c}$ input variables consist of all  \BDT2 input variables, $q^2_{\textrm{rec}}$, and the output of \BDT2.
The BDT$_\mathrm{c}$ classifier is trained with events that satisfy \BDT1$>0.9$ and $\eta(\BDT2)>0.75$ in the off-resonance data and a 50\invfb sample of off-resonance simulation. As a check, BDT$_\mathrm{c}$ is trained using a 200\invfb simulated sample of continuum events produced at a c.m.\ energy corresponding to the \Y4S resonance, yielding a similar performance. 
For the HTA, the BDT$_\mathrm{c}$ exploits all BDTh input variables
and is trained with the off-resonance data and a 1 \invab simulated sample of continuum events produced at a c.m.\ energy corresponding to the \Y4S resonance.
If $p$, taking values between 0.0 and 1.0, denotes the BDT$_\mathrm{c}$ classifier output for a given continuum event, the ratio $p/(1-p)$ approximates
the likelihood ratio $\mathcal{L}(\mathrm{data})/\mathcal{L}(\mathrm{simulation})$, where $\mathcal{L}(\mathrm{data})$ $\left(\mathcal{L}(\mathrm{simulation})\right)$ is the likelihood of the continuum event being from data (simulation), which is used as a weight~\cite{Martschei_2012}. The weights range between 0.5 and 2.0 with a standard deviation of 0.3 for the ITA. This weight, for the ITA, is applied to the simulated continuum events after the final selection; for the HTA, it is applied before the BDTh training.
Comparison of simulated continuum events with off-resonance data shows that the application of this weight improves the modeling of the input variables.

\begin{figure}[htp]
    \centering
    \includegraphics[width=\linewidth]{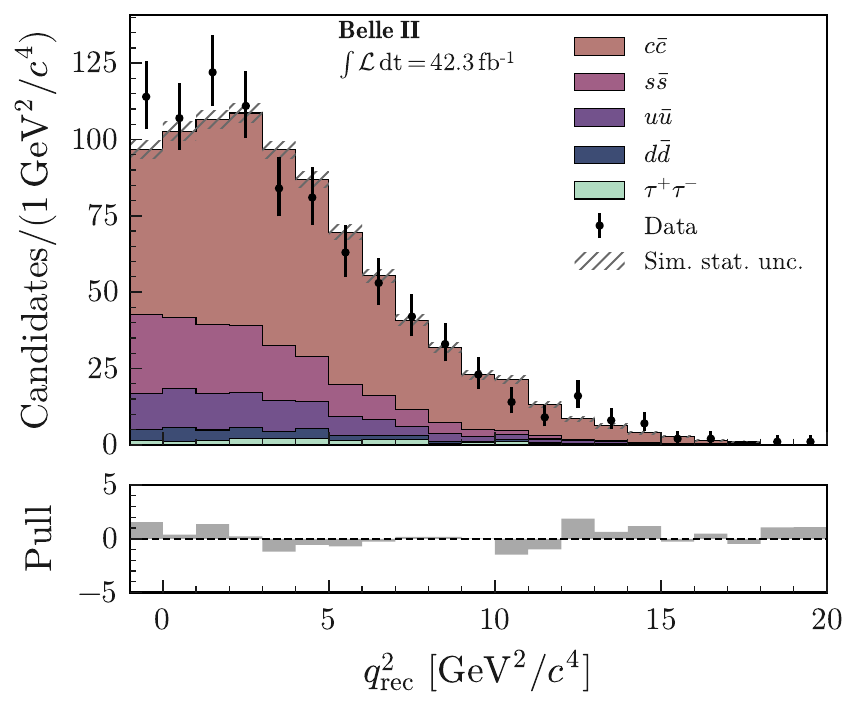}
    \caption{Distribution of $q^2_\mathrm{rec}$ for the off-resonance data (points with error bars) and continuum background simulation (filled histograms) in the SR for the ITA. The simulation is normalized to the number of events in the data.
    The distribution of the difference between data and simulation divided by the combined uncertainty (pull) is shown in the bottom panel.
    }
    \label{fig:q2cont}
\end{figure}
Figure \ref{fig:q2cont} shows a comparison of the $q^2_\mathrm{rec}$ distribution in data and corrected simulation for the ITA off-resonance sample. While the shapes of the distributions are similar, there is a normalization excess of the data over the simulation of
$(40 \pm 5)\%$, which is included as a systematic uncertainty (see \cref{sec:syst}). A possible source of the discrepancy is a mismodeling of kaon fragmentation in the \texttt{PYTHIA8} version used in Belle~II. Illustration of improvement of the ITA distributions with BDT$_\mathrm{c}$-based reweighting is shown in the Supplemental Material~\cite{supplemental}.

For the HTA, the relative normalization between off-resonance data and continuum simulation is  $0.82 \pm 0.01$
before the BDTh selection. This factor accounts for mismodeling effects on the FEI performance for continuum events and is used to scale the expected continuum contamination. 
The relative normalization in the BDTh signal region is consistent with unity with 50\% uncertainty,
which is included as a systematic uncertainty (see \cref{sec:syst}).

\subsection{$\boldsymbol{\B}$ background}
\label{sec:BBbar}
The backgrounds originating from $\B^0$ and $\B^+$ decays are dominant in the highest-sensitivity regions of the analysis.
The composition of the $\B$ backgrounds is similar for both the ITA and HTA samples. It is also similar for \Bp and \Bz decays; however, the contribution from \Bp decays has a larger impact for both analyses.

In the ITA sample, the main background process consists of semileptonic $B$ decays to charm,  where the signal-candidate kaons originate from charmed-meson decays. This process is approximately $47\%$ of the total $\B$ background in the SR. The other major background processes are hadronic $B$ decays involving charmed mesons and other hadronic $B$ decays, contributing about $38\%$ and $14\%$ to the total $\B$ background in the SR, respectively.
The remaining sources of background are 
$\B^+ \to \tau^+ \nu_{\tau}$
decays 
and $\B \to \Kstar \nu \bar \nu$ decays.

In the HTA sample, semileptonic $\B$ decays represent the majority of the $\B$ background events, accounting for approximately $62\%$ of the total background. The second most abundant contribution comes from hadronic $B$ decays with 
final states including a charmed meson accompanied by multiple pions, representing about $20\%$ of the total background. The remaining contributions are from other hadronic modes.

The lower-particle-multiplicity events involving the direct decay of a \B meson into a $D$ meson contribute
more than those containing $D^*$ resonances. The decays involving higher excitations of $D$ mesons ($D^{**}$ modes), which are less well known, correspond to approximately $4\%$
of the total \B background for ITA and $6\%$ for HTA, and are modeled according to their \texttt{PYTHIA8}~\cite{Sjostrand:2014zea} simulation. 
In the following, the modeling of the main background categories and of specific background decays requiring special treatment is presented.

\subsubsection{Modeling of $\D$-meson decays involving a $\Kp$ meson}\label{sec:dkx}
The dominant background contributions in which the signal-candidate \Kp originates from \Dz and \Dp decays are suppressed using several variables that exploit characteristic features of these decays, such as displaced decay vertex and invariant-mass information, as discussed in \cref{sec:bdt}. The modeling of this background is checked by comparing the distributions of these variables in data and simulation at various selection stages, and good agreement is observed.
\begin{figure}[t]
    \centering
    \includegraphics[width=\linewidth]{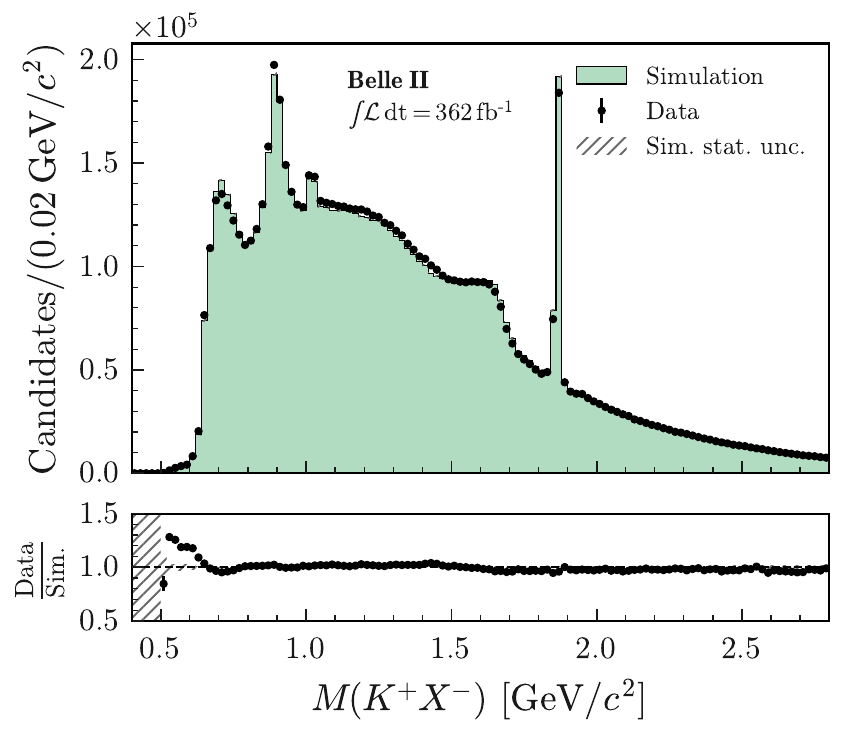}
    \caption{Distribution of invariant mass for the candidate $K^+$ plus a charged particle from the ROE, which is reconstructed under the most probable mass hypothesis among a pion, kaon, proton, electron or muon based on the PID information in data (points with error bars) and the simulation (filled histogram). The samples are shown after the $\BDT1>0.9$ selection in the ITA. The data-to-simulation ratio is shown in the bottom panel.}
    \label{fig:mkx}
\end{figure}
An example is presented in \cref{fig:mkx}, which shows the invariant mass distribution of the signal-kaon candidate paired with a charged particle from the ROE after the \BDT1 selection.~The distinctive shape in data, including the peak from the two-body $\Dz \to K^- \pi^+$ decay, is well reproduced by the simulation.

Uncertainties related to the knowledge of the semileptonic $B$-decay branching fractions are included explicitly, as discussed in \cref{sec:syst}. Uncertainties due to the decay form factors are studied using the 
\texttt{eFFORT} computer program~\cite{markus_prim_2020_3965699} and found to be negligible.

\subsubsection{Modeling of $D$-meson decays involving a \KL meson}\label{sec:dkl}
Backgrounds from prompt production of $K^+$ mesons in \B decays are important in the highest-sensitivity region.
The branching fractions of 
$B^0 \to K^+D^{(\ast) -}$ and $B^+\to K^+\overline{D}{}^{(*)0}$ decays
are relevant due to the sizable and poorly known fraction of $D$-meson decays involving $\KL$ mesons. The branching fraction of decays which involve $B \to D\to \KL$ transitions is studied 
using independent control samples based on alternative particle-identification requirements.
A pion-enriched control sample is used to determine corrections, while samples with the signal track identified as an electron or a muon are used to validate them.

\begin{figure*}
    \centering
    \includegraphics[width=0.49\linewidth]{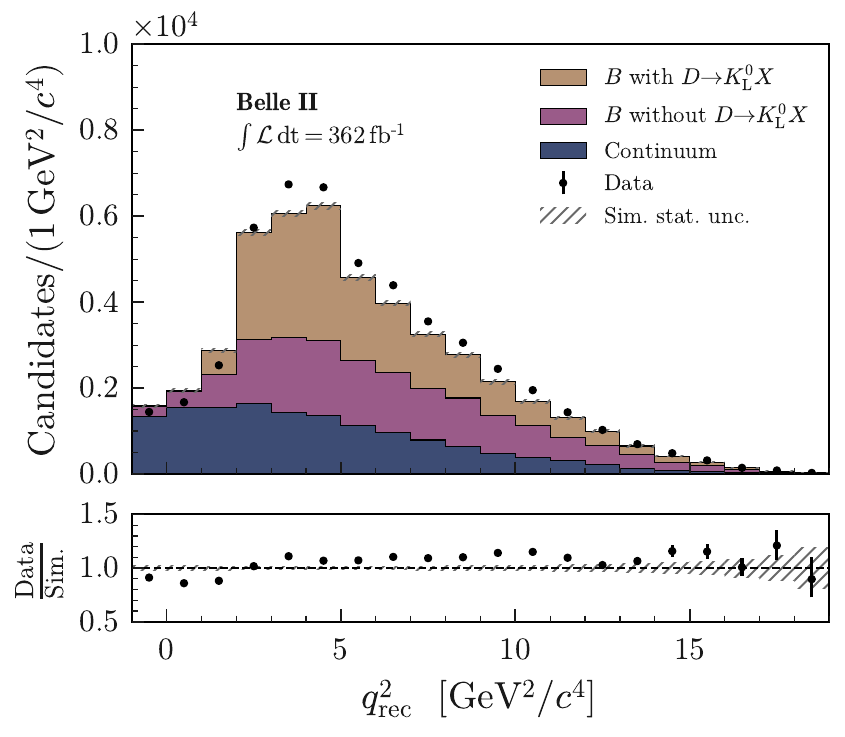}
    \includegraphics[width=0.49\linewidth]{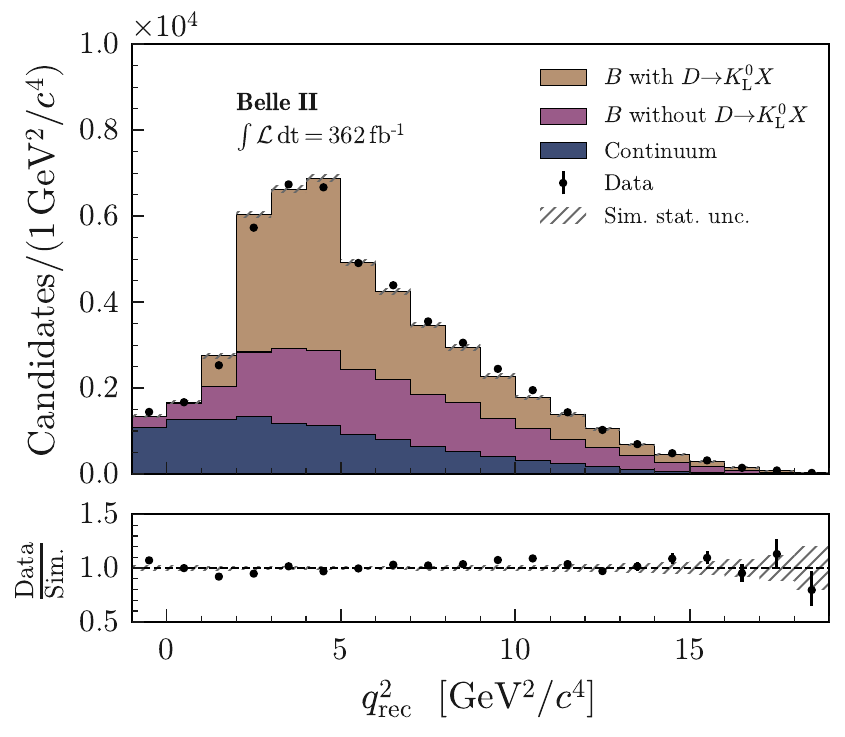}
    \caption{Distribution of $q^2_{\mathrm{rec}}$ in data (points with error bars) and simulation (filled histograms) divided into three groups ($B$-meson decays with and without subsequent $D\rightarrow \KL X$ decays,
and the sum of the five continuum categories) for the pion-enriched sample in the ITA. The left (right) panel shows pre(post)fit distributions. The data-to-simulation ratios are shown in the bottom panels.
    }
    \label{fig:pionSBfits}
\end{figure*}

The pion-enriched sample presents an overall excess of the data over expectations for both ITA and HTA. For the ITA, the excess is studied as a function of $q^2_{\mathrm{rec}}$ and found to appear at the \Dz threshold and above (see \cref{fig:pionSBfits} left). If attributed to $D$-meson decays involving \KL, the excess is consistent with a $(30 \pm 2)\%$ increase in rate compared to the expectation from simulation. This is determined in a three-parameter fit to the binned $q^2_{\mathrm{rec}}$ distribution for $\eta(\BDT2)>0.92$ where the fit parameters are the fractions of summed continuum, summed charged and neutral $B$-meson decays with $D$-meson decays involving \KL, and summed charged and neutral $B$-meson decays without $D$-meson decays involving \KL mesons (see \cref{sec:pionfits} for more details).
 In background simulation the branching fraction for $D^{0}\rightarrow (\Kz/\Kzb)X$ is 40\% and for $D^{+}\rightarrow (\Kz/\Kzb) X$ is 58\%. When these branching fractions are scaled by $1.30$,
 the resulting branching fraction of 52\% for $D^{0}$ is compatible with the known value of $(47 \pm 4)\%$~\cite{ParticleDataGroup:2022pth}; the value for $D^{+}$, 75\%, is above the known value of $(61\pm5)\%$~\cite{ParticleDataGroup:2022pth}. 
 The distribution of $q^2_{\mathrm{rec}}$ in simulation for $B$-meson decays with subsequent  $D^+\to \KL X$ and $D^0\to \KL X$ decays in the pion-enriched ITA control sample is shown in the Supplemental Material~\cite{supplemental}.

An excess at $q^2_{\mathrm{rec}}$ above the charm-production threshold is also evident in the samples 
in which the signal track is identified to be a muon or an electron.
It is covered by $(35 \pm 1)\%$ and $(38 \pm 1)\%$ increases in the rate of charm decays involving $\KL$ in the respective samples. 

Consequently, a correction of $+30\%$ is applied to the branching fraction of events containing $D\to \KL X$ in the simulated background sample, in both ITA and HTA.~The correction is based on the excess size determined for the pion-enriched sample, as the rate of pion-to-kaon misidentification is significantly larger than that of lepton-to-kaon misidentification. Due to the discrepancy in the correction factors between the different samples, a systematic uncertainty of 10\% is assigned; i.e., the correction is (+30 $\pm$ 10)\%.

\begin{figure}[htb]
    \centering
    \includegraphics[width=\linewidth]{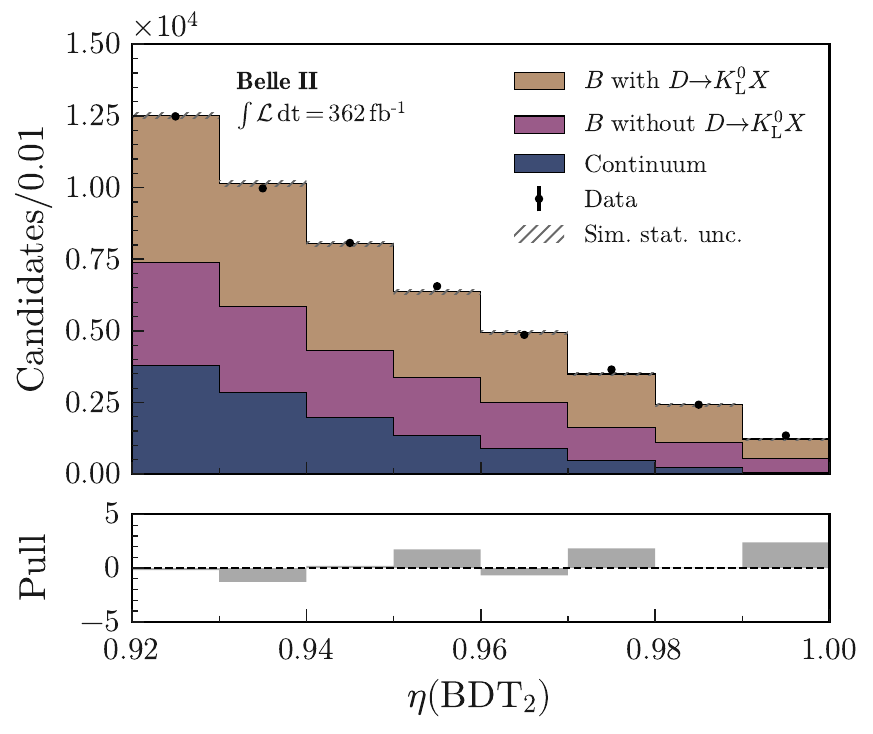}
    \caption{ Distribution of $\eta(\BDT2)$ in data (points with error bars) and simulation (filled histograms) divided into three groups ($B$-meson decays with and without subsequent $D\rightarrow \KL X$ decays, and the sum of the five continuum categories), for the pion-enriched ITA control sample. 
    All the corrections are applied, including the one for the contribution involving $D$ mesons decaying to $\KL$. The pull distribution is shown in the bottom panel.}
    \label{fig:pionbdt}
\end{figure}
\begin{figure*}
    \centering
    \includegraphics[width=0.49\linewidth]{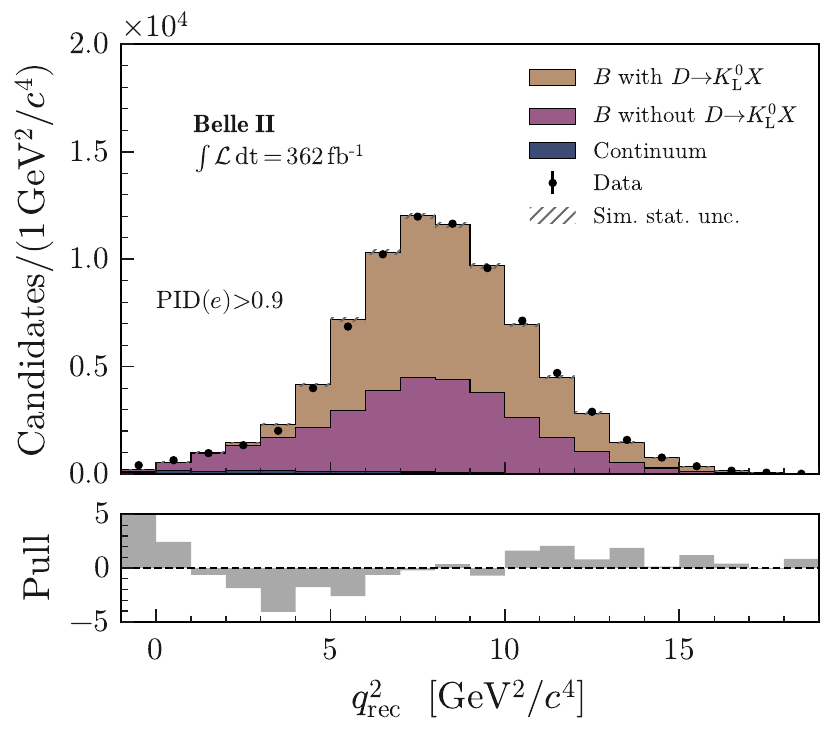}
    \includegraphics[width=0.49\linewidth]{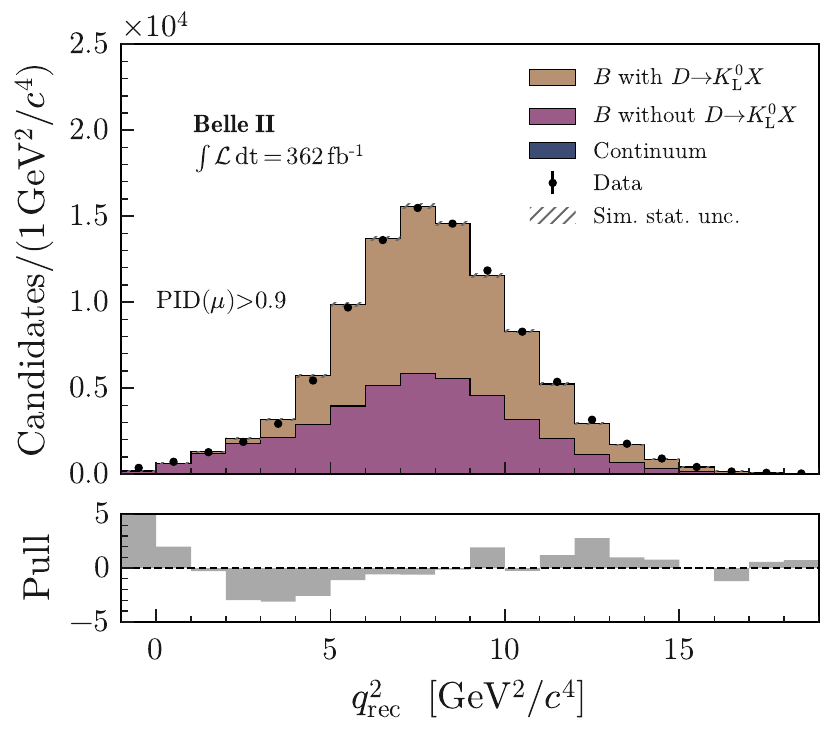}
    \caption{Distribution of $q^2_{\mathrm{rec}}$ in data (points with error bars) and simulation (filled histograms) divided into three groups ($B$-meson decays with and without subsequent $D\rightarrow \KL X$ decays, and the sum of the five continuum categories), 
    for the electron- (left) and muon-enriched (right) PID control samples with $\eta(\BDT2)>0.92$ in the ITA. The pull distributions are shown in the bottom panels.}
    \label{fig:leptonID}
\end{figure*}

Figure \ref{fig:pionbdt}
shows the $\eta(\BDT2)$ distribution for the pion-enriched sample, after all corrections are applied, including the scaling of the branching fraction of $D$-meson decays involving \KL mesons. The resulting expectations are consistent with the data. The $q^{2}_{\mathrm{rec}}$ distribution for the sample is also discussed in \cref{sec:xcheck}. 
The $q^{2}_{\mathrm{rec}}$ distribution for the sample in which the signal track is identified as a lepton is shown in \cref{fig:leptonID}.

\subsubsection[Modeling of background with KL and n]{Modeling of $B^+\to \Kp\Kz\Kzb$, $\Bp \to \Kp n\bar{n}$, $B\to K^{*} \nu \bar{\nu}$, and $B^+\to \tau^+(\to K^+\bar{\nu})\nu$ backgrounds}
\label{sec:btokpklkl}
Another important class of background is charmless hadronic $B$ decays with $\KL$ mesons
or neutrons in the final state, since these neutral particles can mimic
the signal signature. The contributions from $B^+\to \Kp\Kz\Kzb$, $\Bp \to \Kp n\bar{n}$, and $B\to K^{*} \nu \bar{\nu}$ decays are estimated separately, as described in~\cref{sec:data}.
\begin{figure}
\includegraphics[width=\linewidth]{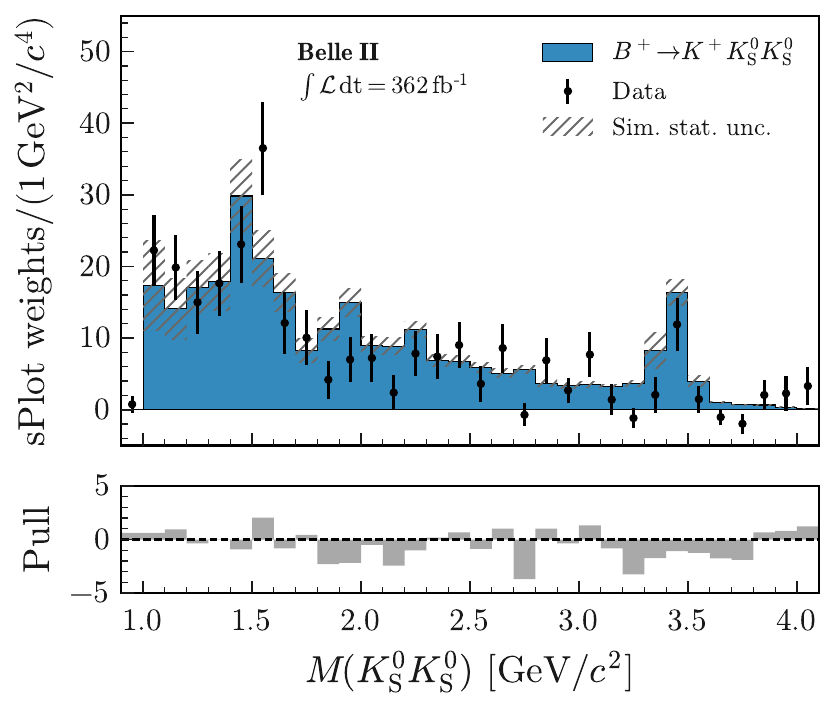}
\caption{Distribution of invariant $\KS\KS$ mass in background-subtracted data (points with error bars) and signal simulation (filled histogram) for $B^+ \to K^+ \KS\KS$ candidates.
The simulated distribution is normalized to the number of $\BBbar$ events.
The pull distribution is shown in the bottom panel.
\label{fig:kpksks} }
\end{figure}

The modeling of the $B^+\to \Kp\Kz\Kzb$ background in the ITA is validated by reconstructing $B^+ \to K^+ \KS\KS$ and $B^0\to \KS K^+K^-$ decays.
Details of the reconstruction are given in~\cref{app:klkl}.
The $s$Plot method~\cite{Pivk:2004ty} is used to determine the invariant-mass distributions for the $\KS\KS$ and $K^+K^-$ pairs.

The result for the $B^+ \to K^+ \KS\KS$ decay is illustrated in \cref{fig:kpksks}. 
Data and simulation show good shape and normalization agreement, validating the $B^+\to K^+ \KL\KL$ modeling.

\begin{figure}[t]
    \centering
    \includegraphics[width=\linewidth]{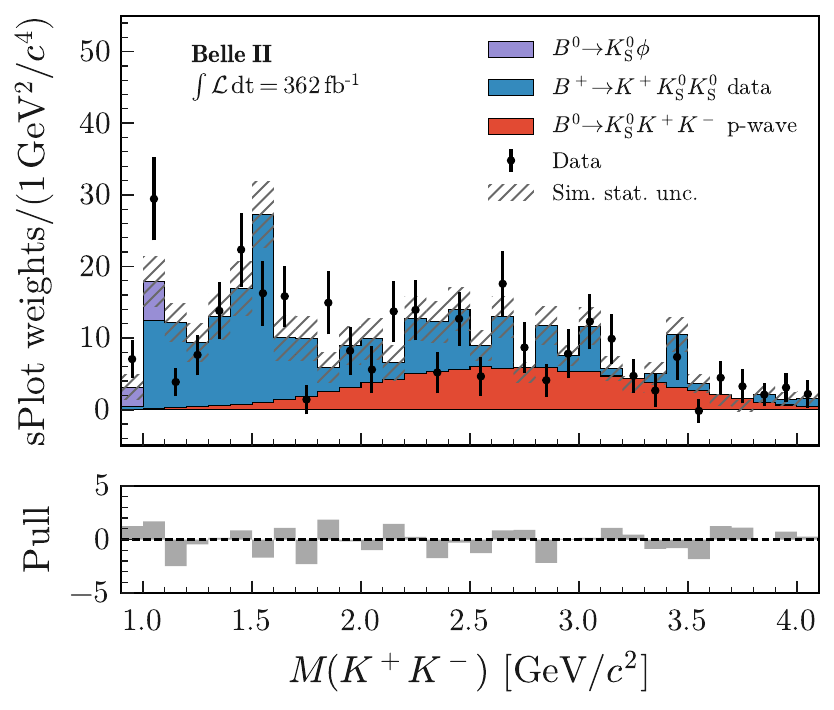}
    \caption{
    Distribution of the invariant mass of the $K^+K^-$ pair from $B^0 \to \KS K^+K^-$ decays in background-subtracted data (points with error bars) and the sum of the simulated
    $B^0 \to \KS \phi(\to K^+K^-)$ decay (purple-filled histogram), the s-wave contribution estimated using  
    $B^+ \to K^+ \KS\KS$ decays in data (blue-filled histogram) and the
    simulated p-wave nonresonant component (red-filled histogram).
    The distribution obtained using $B^+ \to K^+ \KS\KS$ decays in data is corrected for efficiency and the ratio of the $B^{+}$ and $B^{0}$ lifetimes.
    The simulated distributions are normalized to the number of $\BBbar$ events. 
    The pull distribution is shown in the bottom panel.
    }
    \label{fig:kskpkpall}
\end{figure}
The $B^+\to K^+\KL\KS$ decay is modeled as a sum of $B^+ \to \KS \phi$ and
nonresonant p-wave $B^+\to K^+(\KL\KS)_P$ contributions, as described in~\cref{sec:data}. This model is validated by reconstructing the isospin-related decay $B^0\to \KS K^+K^-$ in data.  In addition to $B^0 \to \KS \phi(\to K^+K^-)$ and $B^0\to \KS(K^+K^-)_P$ contributions, this decay proceeds via scalar resonances and a
nonresonant s-wave amplitude. The $B^+ \to K^+ \KS\KS$ decays in data are used to model the latter two contributions only, as this decay lacks a p-wave component due to Bose-Einstein statistics of the $\KS\KS$ pair. Figure \ref{fig:kskpkpall}
shows a comparison between the observed $\KS K^+K^-$ invariant mass and a sum of (1) the $B^+ \to K^+ \KS\KS$ spectrum obtained in data and corrected for efficiency and the ratio of the $B^{+}$ and $B^{0}$ lifetimes, (2) simulated $B^0 \to \KS \phi$ contributions, and (3) simulated $B^0\to \KS(K^+K^-)_P$ contributions. Satisfactory agreement is observed both in shape and normalization.

The $B^{+}\to K^+ n\bar{n}$ background constitutes 0.4\% of the total \B background in the signal region and 1.0\% in the most sensitive region for the ITA. This contribution is significant
because of the threshold enhancement used in the model: these contributions
would be only 0.2\% and 0.3\% if the decay proceeded according to phase
space.

Contaminations from  $B^+\to K^{*+}\nu\bar{\nu}$ and $B^0\to K^{*0}\nu\bar{\nu}$ decays are also included in the background model according to the SM prediction \cite{Parrott:2022zte}. Their expected yield is approximately 5 times smaller than the expected signal yield in the entire SR and 10 times smaller in the most sensitive region.

The long-distance contribution of $B^+\to \tau^+(\to K^+\bar{\nu})\nu$ decay is included as part of the background model (see \cref{sec:data}). Compared to the signal, which by construction has a selection efficiency of 8.0\% in the SR for the ITA, this background has a higher selection efficiency of 9.7\%. This higher efficiency is due to a $q^{2}$ distribution 
that peaks at a lower value than the signal.
However, due to the small branching fraction, the expected yield is approximately 6 times smaller than the expected signal yield in the most sensitive region.

\subsection{Validation of background estimation}
\label{sec:normB}
The modeling of the ITA BDT distributions of background events is tested using events outside the SR with $\eta(\BDT2)$ in the interval 0.75 to 0.90.

For the HTA, the background normalization and BDTh input and output distribution are checked in two control samples: one in which the \Btag and the signal kaon have the same charge and another one in which the requirement on the PID criteria on the signal-side kaon is reversed.

In both analyses, the distributions obtained in data and simulation agree. The normalization of the background contributions also agrees with the expectation.

\section{Signal yield determination} \label{sec:stat}
The signal yields are estimated via binned maximum likelihood fits to
data event counts in the bins of the SRs defined in~\cref{sec:sr}. The ITA
fit is a simultaneous fit to on- and off-resonance data samples; the HTA
fit is to on-resonance data only. 
Templates are used to approximate the distributions, in the relevant observables, of each class of events.
The likelihood function is constructed as a product of Poisson probability-density functions that combine the information from the SR bins.
The systematic uncertainties are included in the likelihood as nuisance parameters, which are approximated as additive or multiplicative modifiers of the relevant yields and constrained to the available auxiliary information using Gaussian likelihoods.
The parameter of interest is $\mu$, the signal branching fraction relative to its SM expectation (signal strength).  
The SM expectation for the signal branching fraction used as a reference is $4.97 \times 10^{-6}$, based on Ref.~\cite{Parrott:2022zte} and excluding the contribution from $\tau$ decays.
The statistical analysis is performed with the \textsc{pyhf} computer program \cite{pyhf,pyhf_joss}, and
the results are checked using a dedicated \textsc{sghf} computer program \cite{Belle-II:2021rof}, which is also used for fits to control samples.

\section{Systematic Uncertainties} \label{sec:syst}
\begin{table*}[]
    \centering
    \caption{Sources of systematic uncertainty in the ITA, corresponding correction factors (if any), their treatment in the fit, their size, and their impact on the uncertainty of the signal strength $\mu$. The uncertainty type can be ``Global'', corresponding to a global normalization factor common to all SR bins, or ``Shape'', corresponding to a bin-dependent uncertainty. Each source is described by one or more nuisance parameters (see the text for more details). The impact on the signal strength uncertainty $\sigma_{\mu}$ is estimated by excluding the source from the minimization and subtracting in quadrature the resulting
      uncertainty from the uncertainty of the nominal fit.}
    \label{tab:systemmatics}
    \begin{ruledtabular}
    \begin{tabular}{lcccc}
      Source &  Correction & Uncertainty  & Uncertainty  & Impact on $\sigma_{\mu}$ \\[-0.1cm]
             &             &    type, parameters      &     size     & \\
        \midrule

      Normalization of $\BBbar$ background &  & Global, 2  & $50\%$ &0.90 \\
      Normalization of continuum background &  & Global, 5  & $50\%$ &0.10 \\
      Leading $B$-decay branching fractions  &   & Shape, 6 & $O(1\%)$ &0.22 \\
      Branching fraction for $B^+\to K^+\KL\KL$ & $q^2$ dependent $O(100\%)$ & Shape, 1  & $20\%$ &0.49 \\
      p-wave component for $B^+\to K^+\KS\KL$ & $q^2$ dependent $O(100\%)$  & Shape, 1  & $30\%$ &0.02 \\
      Branching fraction for $B\to D^{**}$ &  & Shape, 1  & 50\% &0.42 \\
      Branching fraction for $B^+ \to  K^+ n\bar{n}$ & $q^2$ dependent $O(100\%)$ & Shape, 1  & $100\%$ &0.20 \\
      Branching fraction for $D \to \KL X$  &  $+30\%$ & Shape, 1  & 10\% &0.14 \\
      Continuum-background modeling, BDT$_\mathrm{c}$ & Multivariate $O(10\%)$ & Shape, 1  & 100\% of correction &0.01 \\
      Integrated luminosity       &       & Global, 1  & $1\%$ &$<0.01$ \\ 
      Number of $\BBbar$        &       & Global, 1  & $1.5\%$ &0.02 \\
      Off-resonance sample normalization  &  & Global, 1  & 5\% &0.05 \\
      Track-finding efficiency    &       & Shape, 1  & $0.3\%$ &0.20 \\
      Signal-kaon PID             & $p,\theta$ dependent $O$(10--100\%) & Shape, 7  &  $O(1\%)$ &0.07 \\
      Photon energy             &  & Shape, 1  & $0.5\%$ &0.08 \\
      Hadronic energy            & $-10\%$ & Shape, 1  & $10\%$ &0.37 \\
      $\KL$ efficiency in ECL     & $-17\%$ & Shape, 1  & $8.5\%$ &0.22 \\
      Signal SM form factors      & $q^2$ dependent $O(1\%)$& Shape, 3  & $O(1\%)$ &0.02 \\
      Global signal efficiency    &       & Global, 1  & $3\%$ &0.03 \\
      Simulated-sample size               &       & Shape, 156  & $O(1\%)$&~0.52
    \end{tabular}
    \end{ruledtabular}
\end{table*}

A number of possible sources of systematic uncertainty are considered and summarized in~\cref{tab:systemmatics} for the ITA and~\cref{tab:systemmatics_had} for the HTA.

For the ITA, the yields of the seven individual background categories are allowed to vary independently in the fit.
In each case, a Gaussian constraint is added to the fit, centered at the expectation based on (corrected) simulation and with standard deviation corresponding to 50\% of the central value.
The 50\% value is motivated by a global normalization difference between the off-resonance data and continuum simulation, as mentioned in \cref{sec:continuum}. For the charged-\B-background yield, which has the largest correlation with the signal strength $\mu$, the postfit uncertainty is reduced to about half the assigned prefit uncertainty.
The data also significantly constrain the $c\bar{c}$-background yield, reducing the postfit uncertainty to approximately half of the prefit uncertainty.

The remaining systematic uncertainties may also influence the shape of the templates. Each source is described by several nuisance parameters. Several sources are used to cover background-modeling uncertainties. The branching fractions of decay modes contributing about $80\%$ of $B^+$ decays and $70\%$ of $B^0$ decays in the SR are allowed to vary according to their known uncertainties~\cite{ParticleDataGroup:2022pth}.
These variations are then propagated to the SR bins, and their effects, along with correlations, are incorporated into a covariance matrix. This matrix is subsequently factorized into a canonical form using eigendecomposition and represented using six nuisance parameters.
The uncertainty on the branching fraction of the $B^+\to K^+\KL\KL$ decay is estimated to be
$20\%$ to account for potential branching fraction differences 
between  $B^+\to K^+\KL\KL$  and  $B^+\to K^+\KS\KS$ decays. 
The uncertainty on the branching fraction of the $B^+\to K^+\KS\KL$ decay 
 is estimated to be
$30\%$. This accounts for possible isospin-breaking effects ($20\%$) and uncertainties in the p-wave nonresonant contribution ($20\%$).
The uncertainties on the branching fractions of $B\to D^{**}$ decays, which are poorly known,  
are assigned to be $50\%$. 
Uncertainties in the modeling of baryonic decays involving neutrons are covered by the $100\%$ uncertainty on the $B^+ \to K^+n\bar{n}$ branching fraction. The fraction of $D$-meson decays involving $\KL$ mesons is corrected by $30\%$ with a $10\%$ absolute uncertainty, motivated by the differences in the scaling factors determined using different samples, as discussed in \cref{sec:dkl}. All of these uncertainties are propagated as correlated shape uncertainties. 

Global normalization uncertainties on the luminosity measurement ($1\%$ assumed) and the number of $\BBbar$ pairs ($1.5\%$) are treated with one nuisance parameter each. In addition, a $5\%$ uncertainty is introduced on the difference in normalization between on- and off-resonance data samples.

The following five sources represent uncertainties in detector modeling; they are discussed in detail in \cref{sec:corrs}. The sources are track-finding efficiency,  kaon-identification efficiency, modeling of energy for photons and hadrons, and $\KL$ reconstruction efficiency. The final three sources account for signal-modeling uncertainties. These are signal form factors, which are based on Ref.~\cite{Parrott:2022zte}, and global signal-selection efficiency uncertainties as determined in \cref{sec:sigeff}.

The systematic uncertainty due to the limited size of simulated samples is taken into account by one nuisance parameter per bin per category (156 parameters).

To account for all the systematic sources described above, a total of 193 nuisance parameters, along with the signal strength $\mu$, are varied in the fit.

The largest impact on the uncertainty of the signal strength $\mu$ arises from the knowledge of the normalization of the background from charged \B decays. Other important sources are the simulated-sample size, branching fraction for $B^+\to K^+\KL\KL$ decays, branching fraction for $B \to D^{**}$ decays,
reconstructed energy of hadrons, branching fractions of the leading $B$ decays, and $\KL$ reconstruction efficiency.
These sources of uncertainty allow for substantial changes in the $\BBbar$ shape. The shape variations
are larger than the data-simulation residuals in $\eta(\BDT2)$ in
the pion-enriched sample~(\cref{fig:pionbdt}). This suggests that uncertainties in
the $\BBbar$ shape are adequately covered by the existing systematic
contributions. 

\begin{table*}[]
    \caption{Sources of systematic uncertainty in the HTA (see caption of Table \ref{tab:systemmatics} for details).}
  \label{tab:systemmatics_had}
  \centering
\begin{ruledtabular}
\begin{tabular}{lcccc}
Source &  Correction & Uncertainty type,  & Uncertainty size  & Impact on $\sigma_{\mu}$ \\
      &              &  parameters        &                    &                         \\
\midrule
Normalization of $\BBbar$ background &  & Global, 1 & 30\% & 0.91\\
Normalization of continuum background &  & Global, 2  & 50\% & 0.58\\
Leading $B$-decay branching fractions  &   & Shape, 3 & $O(1\%)$ &  0.10\\
Branching fraction for $B^+\to K^+\KL\KL$ & $q^2$ dependent $O(100\%)$ & Shape, 1 & $20\%$ & 0.20\\
Branching fraction for $B\to D^{**}$ &  & Shape, 1  & 50\% & $<0.01$\\
Branching fraction for $B^+ \to K^+n\bar{n}$ & $q^2$ dependent $O(100\%)$ & Shape, 1 & $100\%$ & 0.05\\
Branching fraction for $D \to \KL X$  &  $+30\%$ & Shape, 1 & 10\% & 0.03\\
Continuum-background modeling, BDT$_\mathrm{c}$ & Multivariate $O(10\%)$ & Shape, 1 & 100\% of correction & 0.29\\
Number of $\BBbar$        &       & Global, 1  & $1.5\%$ & 0.07\\
Track finding efficiency    &       & Global, 1  & $0.3\%$ & 0.01\\
Signal-kaon PID             & $p,\theta$ dependent $O$(10--100\%) & Shape, 3 &  $O(1\%)$ & $<0.01$\\
Extra-photon multiplicity & $n_{\gamma\text{extra}}$ dependent $O(20\%)$ & Shape, 1  & $O(20\%)$& 0.61\\
$\KL$ efficiency     &  & Shape, 1  & $17\%$ & 0.31\\
Signal SM form factors      & $q^2$ dependent $O(1\%)$& Shape, 3  & $O(1\%)$ & 0.06\\
Signal efficiency    &       & Shape, 6  & $16\%$ & 0.42\\
Simulated-sample size  &       & Shape, 18 & $O(1\%)$& 0.60\\
\end{tabular}
\end{ruledtabular}
\end{table*}
The summary of systematic uncertainties for the HTA is provided in Table \ref{tab:systemmatics_had}.
Three background components are considered in the HTA: $\BBbar$, accounting for both charged and neutral \B decays; $c\bar c$; and light-quark continuum ($u\bar{u}$, $d\bar{d}$, $s\bar{s}$). The contribution from $\tau$-pair decays is negligible. 
The primary contribution to the systematic uncertainty arises from the determination of the normalization of the $\BBbar$ background. This determination is based on the comparison of data-to-simulation normalization in the pion-enriched control sample, which shows agreement within the 30\% statistical uncertainty.
The other important sources are the uncertainty associated with the bin-by-bin correction of the extra-photon-candidate  multiplicity, and the uncertainty due to the limited size of the simulated sample. The uncertainty on continuum normalization (50\%), determined using off-resonance data, is the fourth most important contribution.
The limited size of the HTA sample prevents the substantial reduction of postfit uncertainties seen in the ITA, compared to prefit values, for the background normalization. 
The other sources of systematic uncertainty are the same in both analyses, except for those related to photon and hadronic-energy corrections, not applied in the HTA, and the p-wave contribution from $B^+\to K^+\KS \KL$, whose contribution is negligible.

For both analyses, nuisance-parameter results are investigated in detail.
No significant shift is observed for the background yields from charged and neutral \B-meson decays.
For the ITA, the shifts in the continuum background yields are consistent with the difference observed in the normalization of the continuum simulation with respect to the off-resonance data.

\section{Results} \label{sec:result}

\begin{figure*}[htp]
\centering
\includegraphics[width=0.49\linewidth]{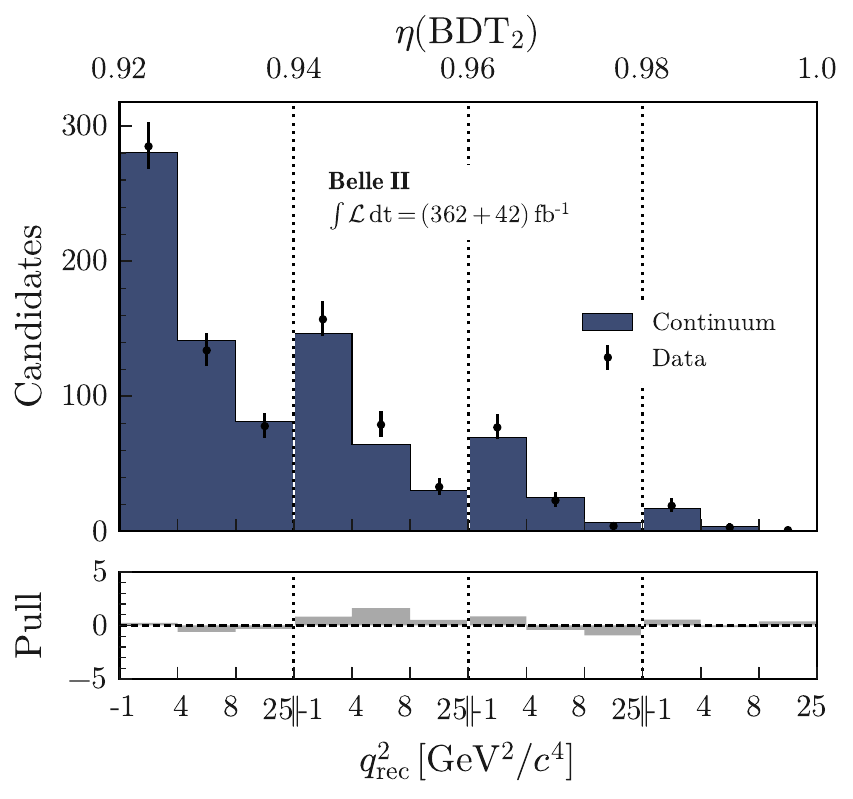}
\includegraphics[width=0.49\linewidth]{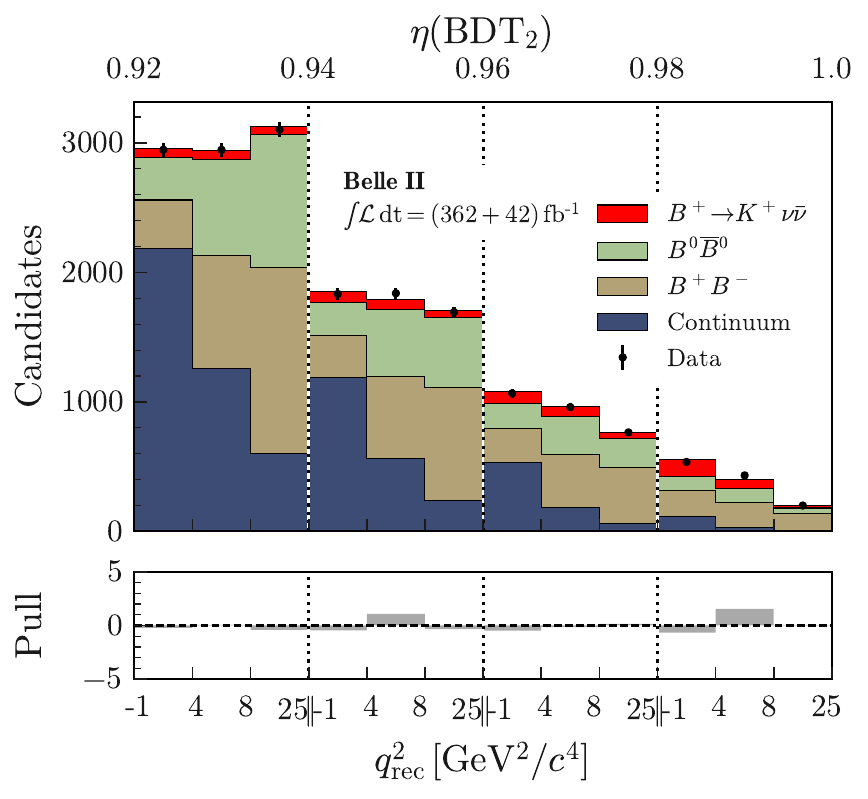}
\caption{Observed yields and fit results in bins of the $\eta(\BDT2)\times q^2_{\mathrm{rec}}$ space obtained by the ITA simultaneous fit to the off- and on-resonance data,
  corresponding to an integrated luminosity of 42 and 362\invfb, respectively.
	The yields are shown individually for the \BKnn  signal, neutral and charged \B-meson decays and the sum of the five continuum categories. 
    The yields are obtained in bins of the $\eta(\BDT2)\times q^2_{\mathrm{rec}}$ space.
    The pull distributions are shown in the bottom panel.
	}
\label{fig:yields}
\end{figure*}

\begin{figure}
    \centering
\includegraphics[width=\linewidth]{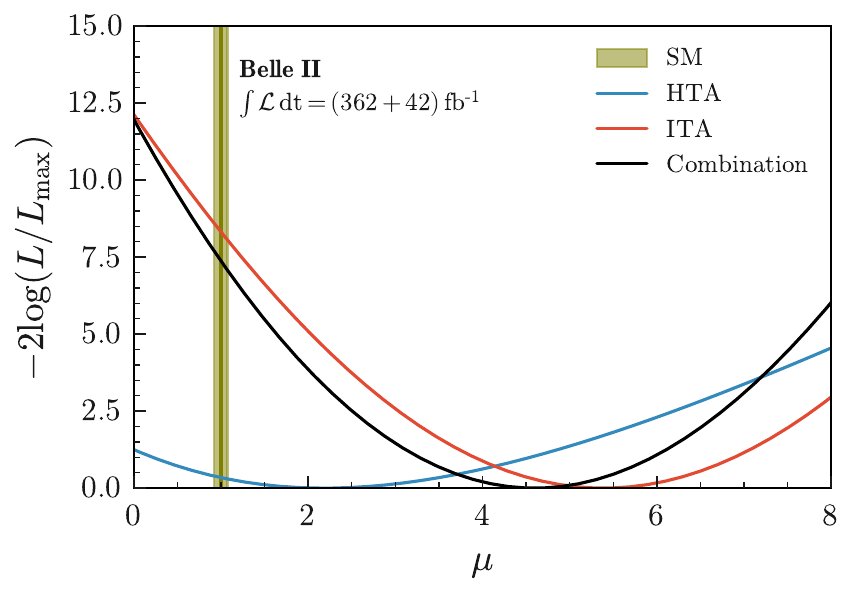}
    \caption{Twice the negative profile log-likelihood ratio as a function of the signal strength $\mu$ for the ITA, HTA, and the combined result. The value for each scan point is determined by fitting the data, where all parameters but $\mu$ are varied.}
    \label{fig:scan} 
\end{figure}

\begin{figure*}[htp]
    \centering
    \includegraphics[width=0.49\linewidth]{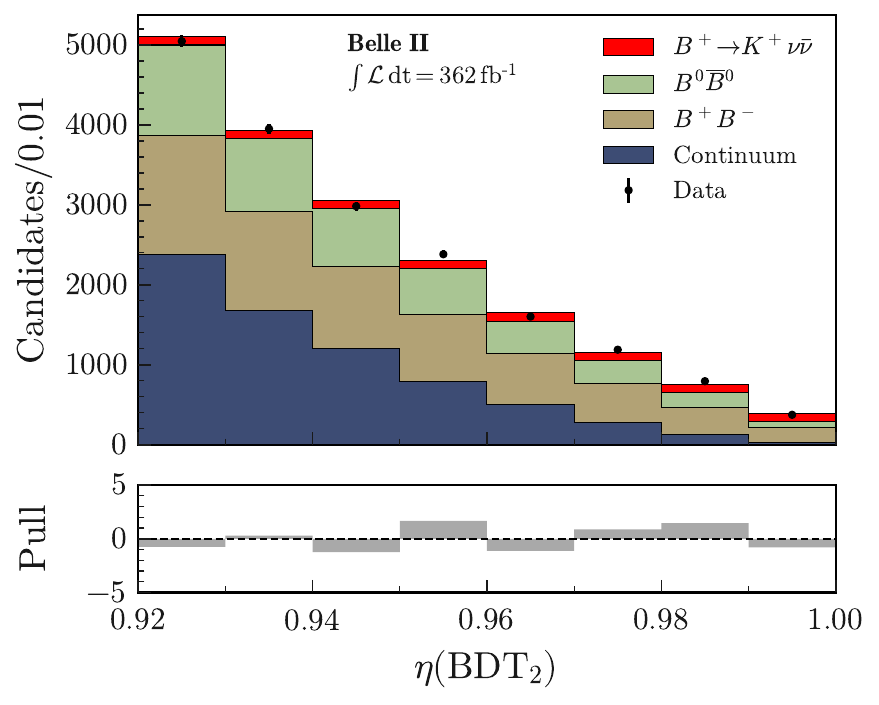}
    \includegraphics[width=0.49\linewidth]{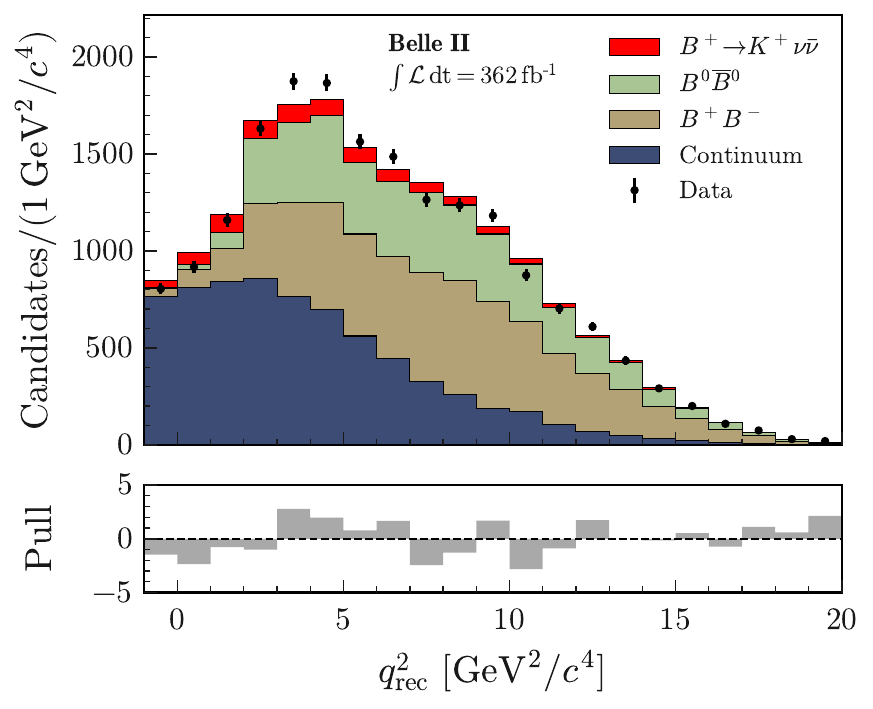}
    \includegraphics[width=0.49\linewidth]{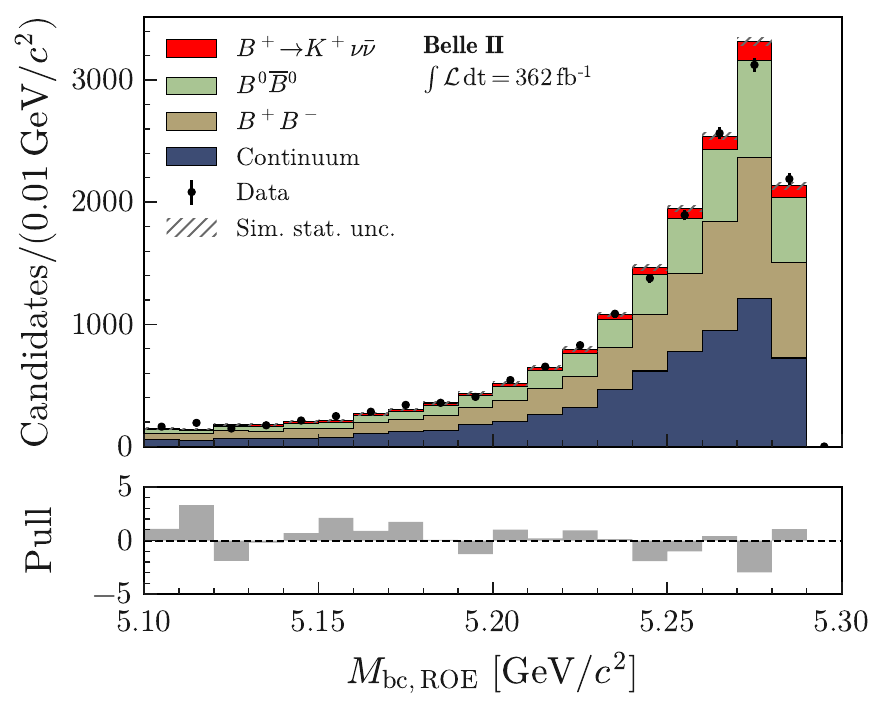}
    \includegraphics[width=0.49\linewidth]{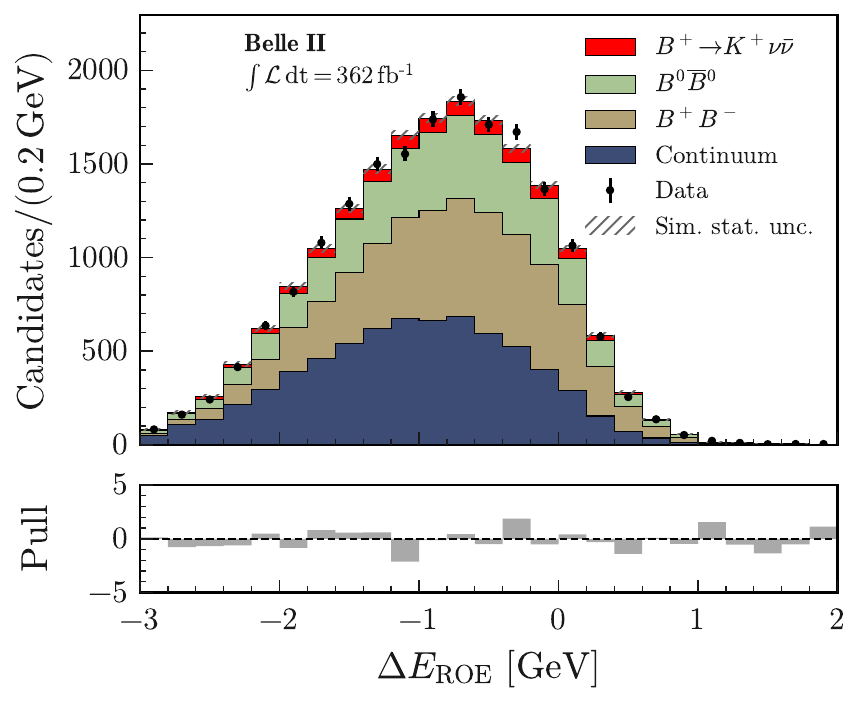}
    \includegraphics[width=0.49\linewidth]{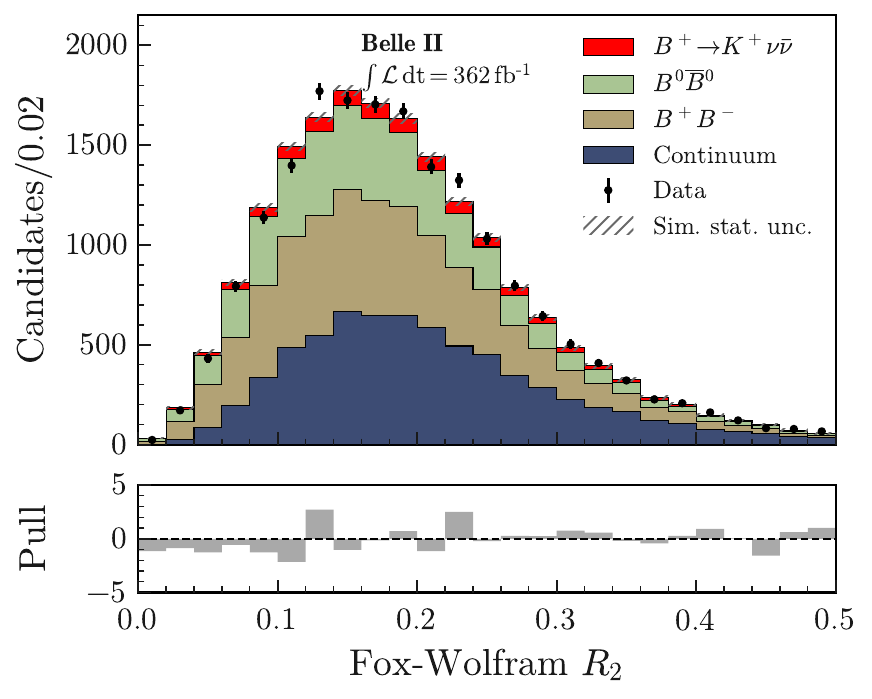}
    \includegraphics[width=0.49\linewidth]{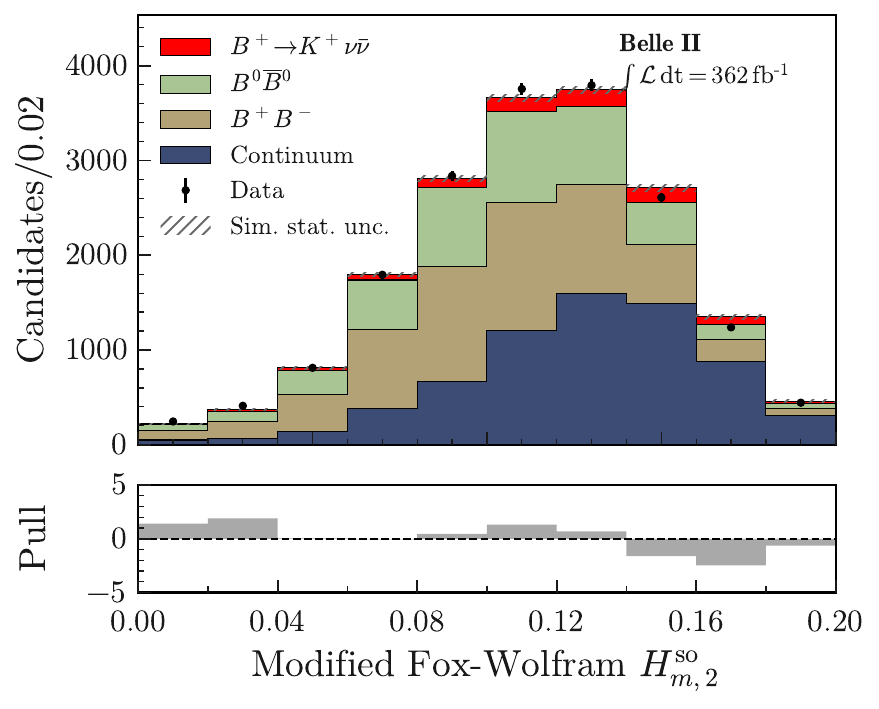}
\caption{Distributions of $\mathit{\eta}$(BDT$_2$), $q^2_{\textrm{rec}}$, beam-constrained mass of the ROE $M_{{\rm{bc},\textrm{ROE}}}$,  $\Delta E_{\textrm{ROE}}$, Fox-Wolfram $R_2$, and modified Fox-Wolfram $H_{m,2}^{\rm so}$ 
in data (points with error bars) and simulation (filled histograms) shown individually for the \BKnn  signal, neutral and charged \B-meson decays, and the sum of the five continuum categories in the ITA. Events in the full signal region, with $\eta(\rm{BDT_{2}}) > 0.92$, are shown. 
Simulated samples are normalized according to the fit yields in the ITA.  
The pull distributions are shown in the bottom panels.}
\label{fig:itafigs_a}
\end{figure*}

\begin{figure*}[htp]
    \centering
    \includegraphics[width=0.49\linewidth]{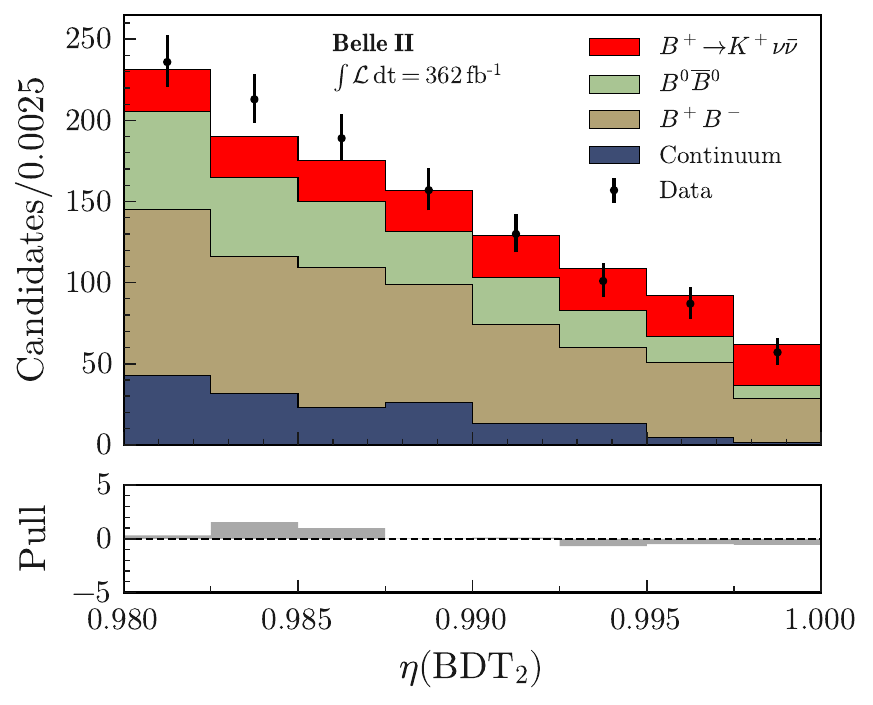}
    \includegraphics[width=0.49\linewidth]{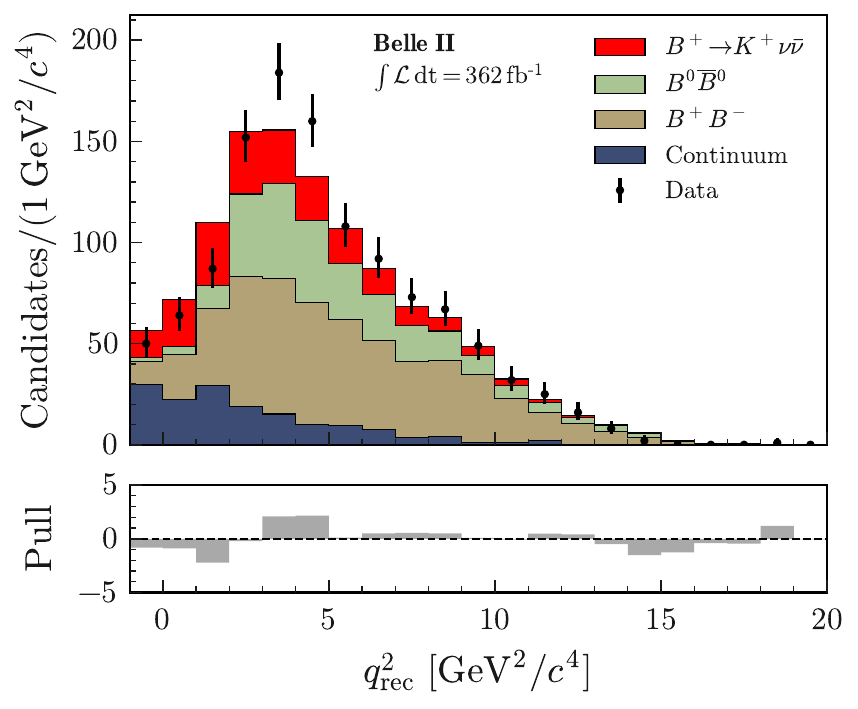}\\
    \includegraphics[width=0.49\linewidth]{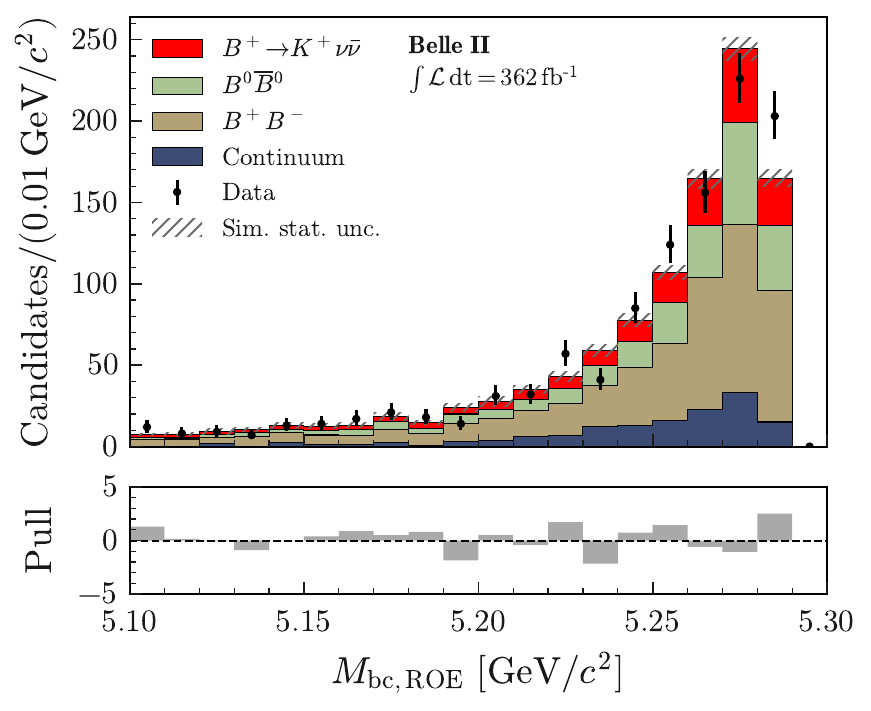}
    \includegraphics[width=0.49\linewidth]{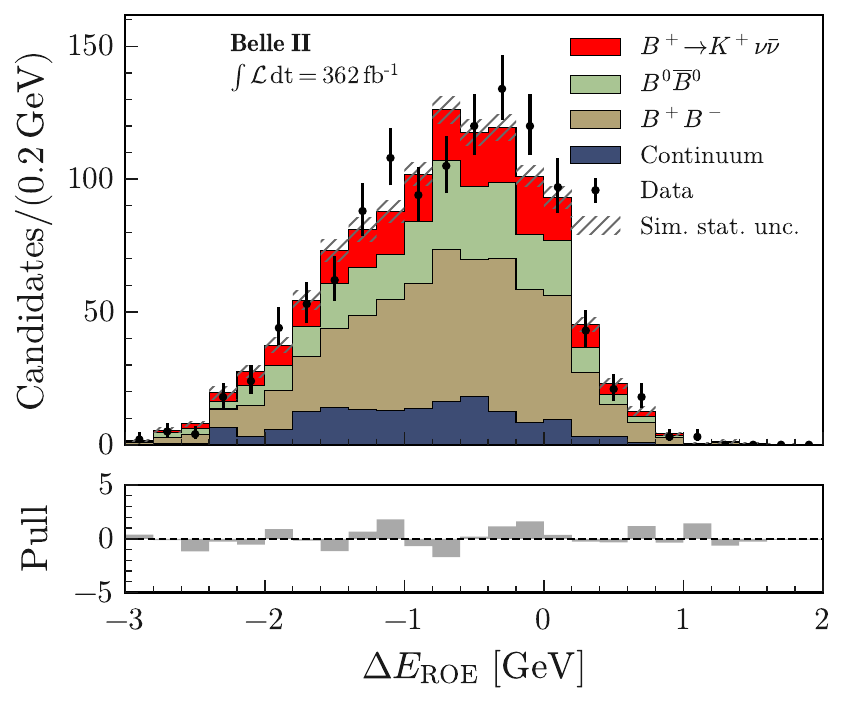}
    \includegraphics[width=0.49\linewidth]{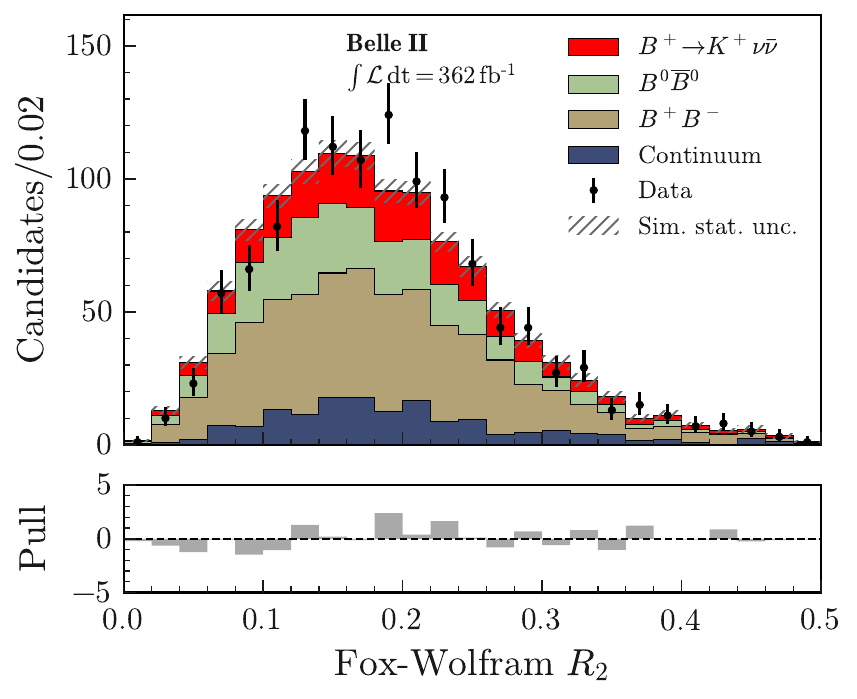}
    \includegraphics[width=0.49\linewidth]{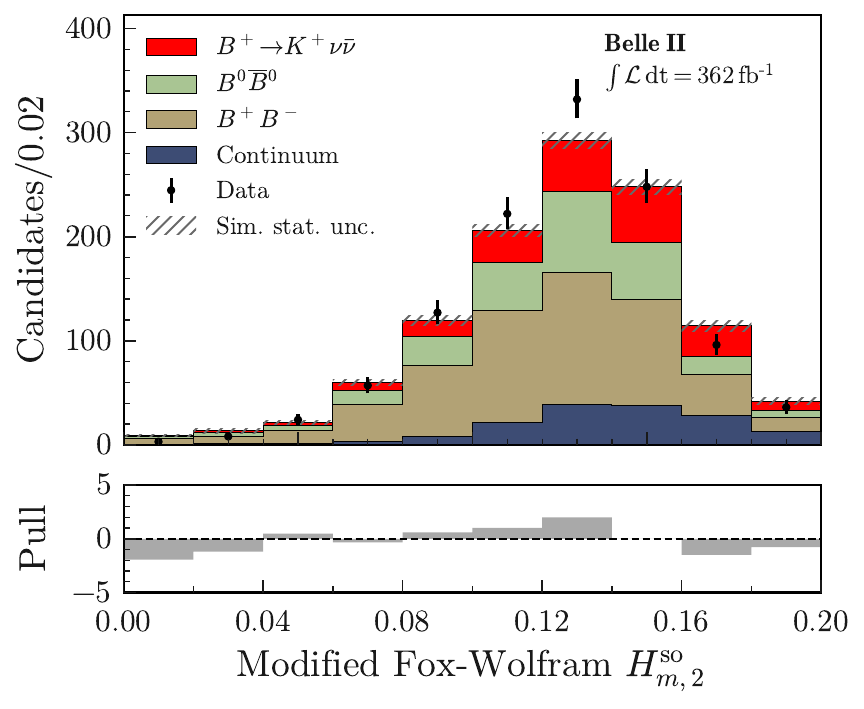}

\caption{Distributions of $\mathit{\eta}$(BDT$_2$), $q^2_{\textrm{rec}}$, beam-constrained mass of the ROE $M_{{\rm bc,\textrm{ROE}}}$, $\Delta E_{\textrm{ROE}}$, Fox-Wolfram $R_2$, and modified Fox-Wolfram $H_{m,2}^{\rm so}$ 
in data (points with error bars) and simulation (filled histograms) shown individually for the \BKnn  signal, neutral and charged \B-meson decays, and the sum of the five continuum categories in the ITA. Events in the most signal-rich region, with $\eta(\rm{BDT_{2}}) > 0.98$, are shown. 
Simulated samples are normalized according to the fit yields in the ITA.  
The pull distributions are shown in the bottom panels.}
\label{fig:itafigs_b}
\end{figure*}

The data in the off-resonance sample and in the SR of the ITA are shown in \cref{fig:yields}, with fit results overlaid. 
Good visual agreement between data and fit is observed in both samples.
An excess over background is observed in the SR, consistent with the presence of \BKnn signal.
The observed signal purity, in terms of the fraction of signal events, is 5\% in the SR and 19\% in the three bins with $\eta(\BDT2)>0.98$.
The signal strength is determined to be  
$
\mu = 5.4\pm 1.0(\mathrm{stat}){}\pm 1.1(\mathrm{syst})= 5.4\pm 1.5,
$
where the statistical uncertainty is estimated using simplified simulated experiments based on Poisson statistics. 
The total uncertainty is obtained using a profile likelihood ratio, fitting the model with fixed values of $\mu$ around the best-fit value while keeping the other fit parameters free; see \cref{fig:scan}.
The systematic uncertainty is calculated by subtracting the statistical uncertainty in quadrature from the total uncertainty. 
An additional 8\% theoretical uncertainty, arising from the knowledge of the branching fraction is not included.
Compatibility between the data and fit result is
assessed using simplified experiments, and a $p$-value of 47\% is found.
(The test is based on the fraction of simplified experiments with the
negative profile log-likelihood ratio of the nominal to the ‘‘saturated’’ model,
in which the predictions are set to the observations, above the one
observed in data.) 
Figures~\ref{fig:itafigs_a} and \ref{fig:itafigs_b}
present distributions of several variables for the events within the signal region. Distributions of $q^{2}_{rec}$ with more differential background information are included in the Supplemental Material~\cite{supplemental}. Each simulated event is weighted using the ratio of post-to-prefit yields for the corresponding SR bin and event category. Good overall agreement is observed. 
However, certain discrepancies are evident in the $q^2_{\mathrm{rec}}$ distribution, showing a deficit in data-to-predictions for $q^2_{\mathrm{rec}}<3 \gevgevcccc$ and an excess for $3 \gevgevcccc<q^2_{\mathrm{rec}}<5\gevgevcccc$.

\begin{figure}[htp]
\centering
\includegraphics[width=\linewidth]{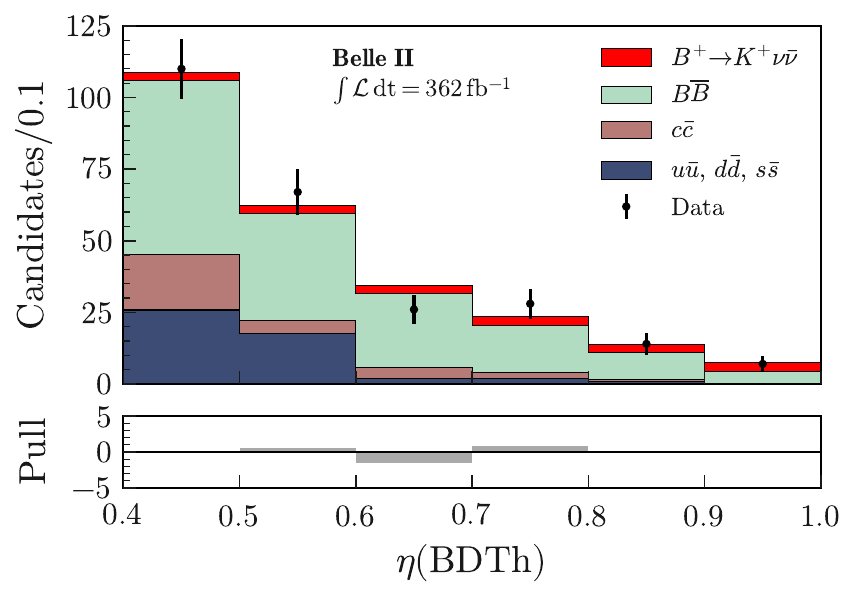}
\caption{\label{fig:yieldsh}Observed yields and fit results in bins of $\eta(\mathrm{BDTh})$ as obtained by the HTA fit, corresponding to an integrated luminosity of 362 \invfb. The yields are shown for the \BKnn signal and the
three background categories ($\BBbar$ decays, $c\bar{c}$ continuum, and light-quark
continuum).
The pull distribution is shown in the bottom panel.
}
\end{figure}

\begin{figure*}[htp]
\centering
\subfloat{\includegraphics[width = 3.5in]{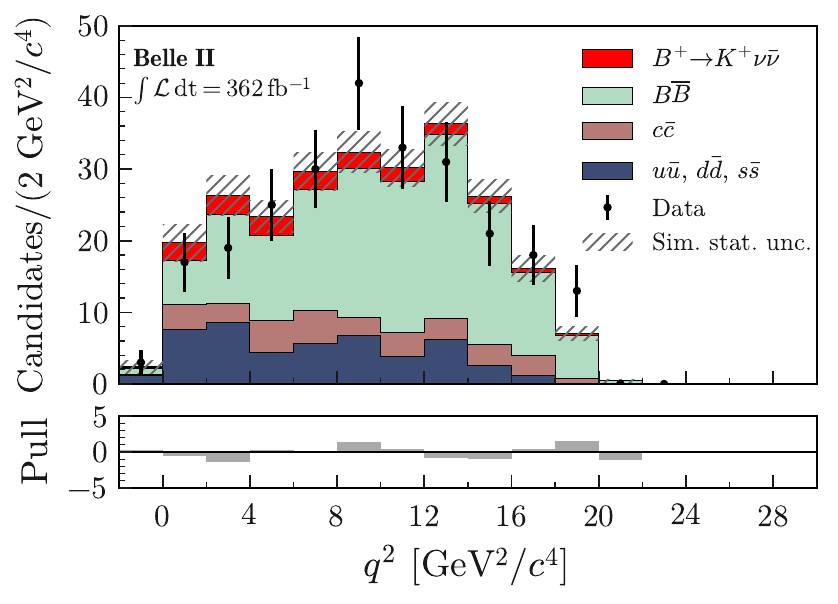}}
\subfloat{\includegraphics[width = 3.5in]{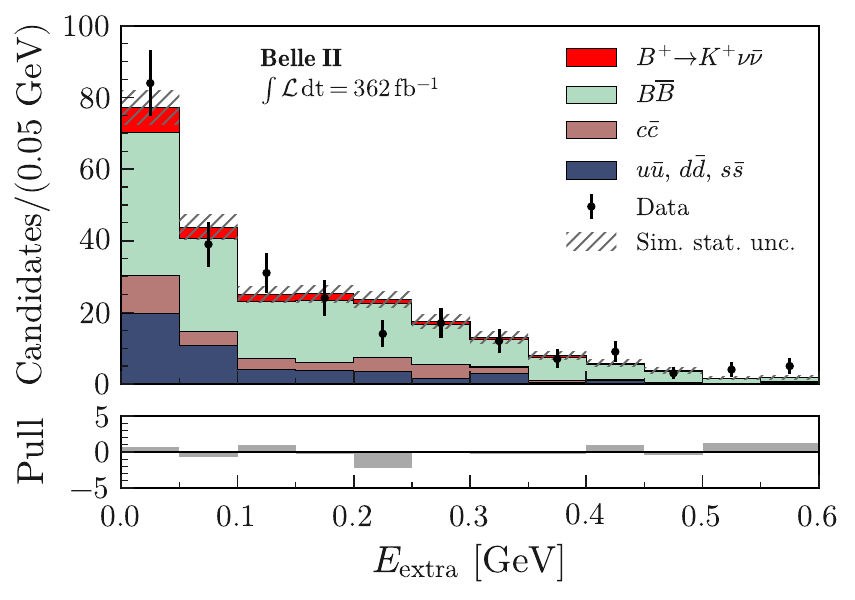}}\\
\subfloat{\includegraphics[width = 3.5in]{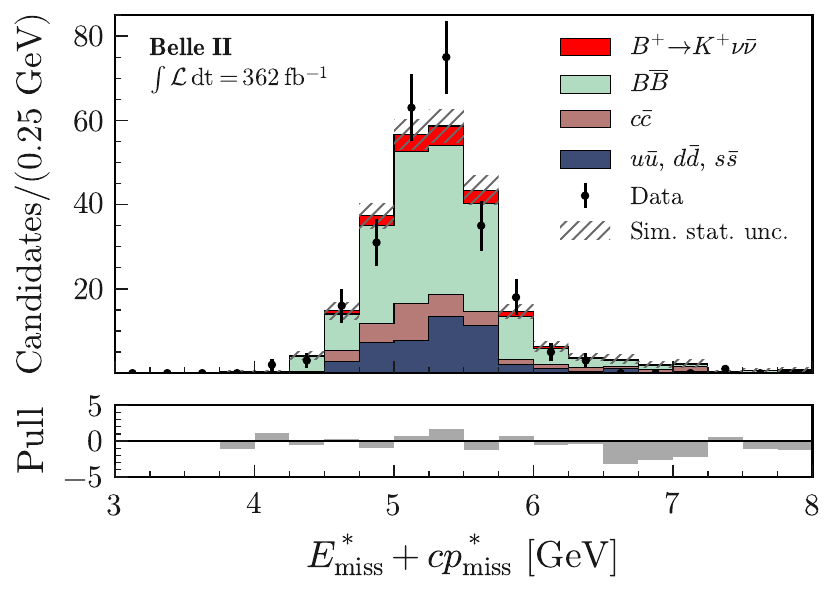}}
\subfloat{\includegraphics[width = 3.5in]{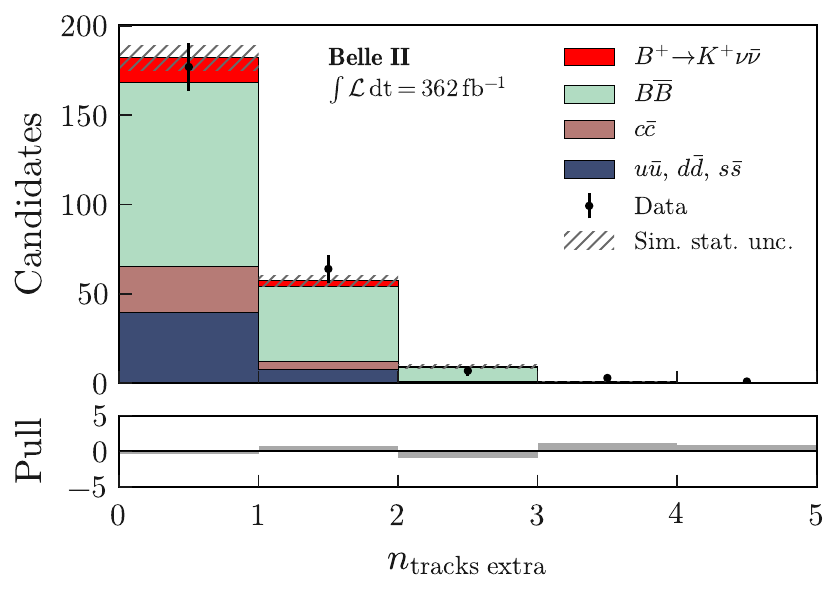}}\\
\caption{Distribution of $q^2$, computed using \Btag kinematics,  \Eecl, \sumEp, and $n_{\rm{tracks~extra}}$ for events in the signal region of the HTA. These distributions are obtained for $B^+ \to K^+ \nu \bar \nu$ candidates reconstructed in data (points with error bars), simulation (filled histograms) of 
 \BKnn signal, $\BBbar$ decays, $c\bar{c}$ continuum, and light-quark continuum backgrounds, normalized according to the fit yields in the HTA.
The pull distributions are shown in the bottom panels.}
\label{fig:distr_postfit_had_1}
\end{figure*}

A comparison of data and fit results for the HTA is shown in \cref{fig:yieldsh}.
The compatibility between the data and fit results is determined to be $61\%$.
The HTA observes a signal strength of $\mu = 2.2^{+1.8}_{-1.7}(\mathrm{stat}){}^{+1.6}_{-1.1}(\mathrm{syst}) = 2.2^{+2.4}_{-2.0}$, lower than the
ITA result.
In the whole SR, a signal purity of 7\% is measured, which increases to $20\%$ in the three bins with $\eta(\mathrm{BDTh})>0.7$, with the main background contribution from 
$\BBbar$ decays.
Figure \ref{fig:distr_postfit_had_1}
shows distributions of several variables for the events within the signal region. Good agreement is observed. Limit setting for HTA is included in the Supplemental Material~\cite{supplemental}.

If interpreted in terms of signal, the results correspond to a branching fraction of the \BKnn decay of
 \ITABFdetailed for the ITA and
 \HTABFdetailed for the HTA.  As mentioned in \cref{sec:stat}, the measured branching fraction does not include the contribution from the long-distance double-charged-current $B^+\to \tau^+(\to K^+\bar{\nu})\nu$ decay.
 
The significance of the observation is determined
by evaluating the profile likelihood $L$ for several $\mu$ values.
The square root of the difference between the $-2\ \textrm{log}\ L$ values at $\mu=0$ and the minimum is used to estimate the significance of the observed excess with respect to the background-only hypothesis,
which yields \ITAsigO standard deviations for the ITA. For the HTA, the observed signal is consistent with the background-only hypothesis at \HTAsigO standard deviations.
Similarly, the square root of the difference between the  $-2\ \textrm{log}\ L$ values at $\mu=1$ and at the minimum is used to estimate the significance of the observed signal with respect to the SM expectation. For the ITA, it is found to be \ITAsigSM standard deviations, indicating a potential deviation from the SM. For the HTA, the result is in agreement with the SM at \HTAsigSM standard deviations.

Events from the SR of the HTA represent only 2\% of the corresponding events in the ITA; their removal
does not alter the ITA result significantly. The ITA sample with removed overlapping events is used for the compatibility checks.
The ITA and HTA measurements agree, with a difference in signal strength of $1.2$ standard deviations.

\section{Consistency checks} \label{sec:xcheck}
Several checks are performed to test the validity of the analysis.

\begin{figure*}
    \centering
\includegraphics[width=0.49\linewidth]{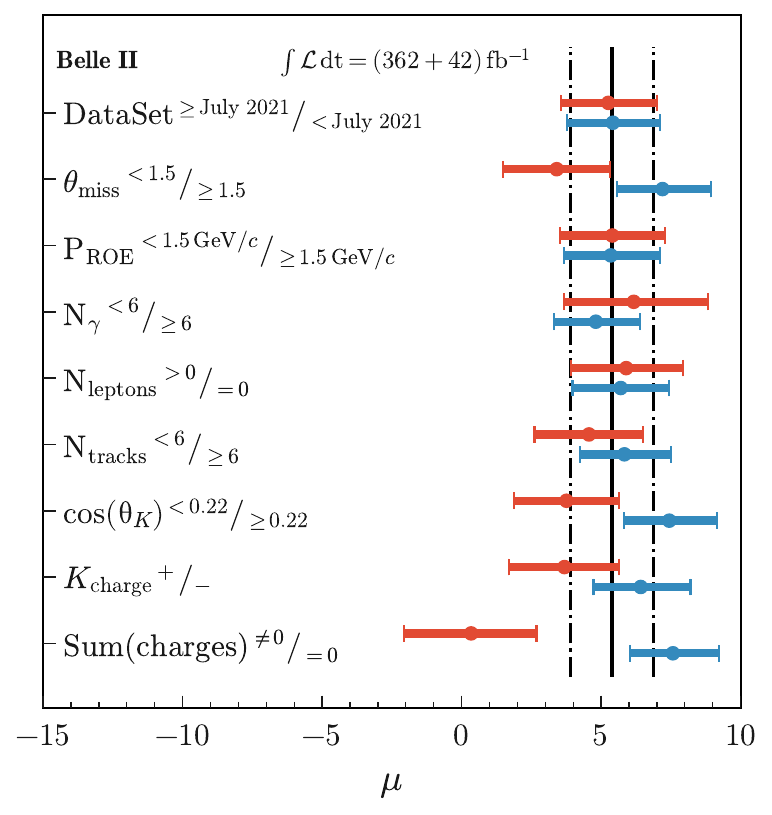}
\includegraphics[width=0.49\linewidth]{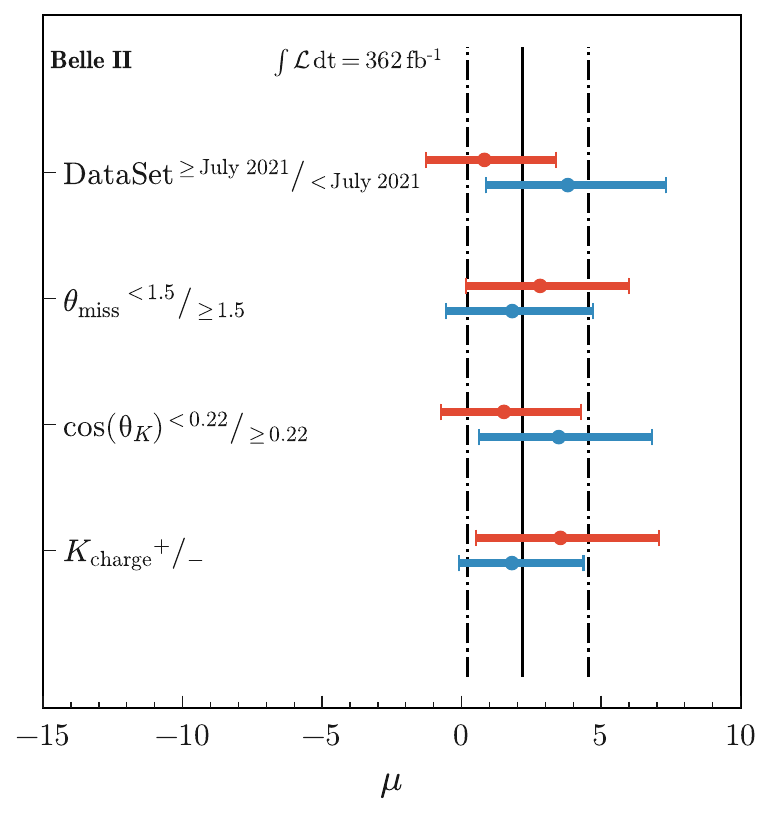}
    \caption{Signal strength $\mu$ determined in the ITA (left) and HTA (right) for independent data samples divided into approximate halves by various criteria. The vertical lines show the result obtained on the full data set. The horizontal bars (and
dot-dashed lines) represent total 1 standard deviation uncertainties.}
    \label{fig:split}
\end{figure*}
Simulation and data events are divided into approximately same-size statistically independent samples (split samples) according to various criteria: data-taking period; missing-momentum direction; momentum of the rest-of-event particles; number of photons, charged particles, and lepton candidates in the event; kaon direction; kaon charge; and total charge of the reconstructed particles in the event. Fits are performed for each split sample, and the results are presented in~\cref{fig:split}.

Good compatibility is observed between the split samples for the HTA. A tension at 2.4 standard deviations is observed for the total charge split sample in the ITA. Several studies are conducted to investigate this tension, but they did not reveal any significant systematic effects. The total $\chi^2$ value per degrees of freedom for all tests in the ITA is $12.5/9$.

An important test involves the subdivision based on the number of leptons in the ITA. Since there are no leptons on the signal side, this test compares events in which a (semi)leptonic $B$ decay occurs in the ROE with those in which a hadronic $B$ decay occurs.
The separation is confirmed by inspecting simulated events. Excellent agreement is observed between the results in the two split samples. This demonstrates the robustness of the ITA procedure with respect to a particular signature in the ROE.

For each common split sample, a comparison is also performed between the ITA and the HTA, showing compatibility between 1 and 2 standard deviations.

An ITA fit fixing the normalization of the \B background to the expectation and the normalization of the continuum to the yield observed in off-resonance data yields a reduction of the uncertainty on $\mu$ by $25\%$ with a downward change in $\mu$
that is consistent with zero at $1.5$ standard deviations. Performing a fit where the 50\% constraints on the normalizations of all background sources are released leads to a minimal change of $\mu$ by $0.1$, with the uncertainty on $\mu$ increasing by only 5\%.
Another fit in which the leading systematic uncertainties are fixed also gives a consistent result.
 A fit to the 12 bins of \Y4S data only, i.e., excluding the off-resonance data, changes $\mu$ by less than $0.1$, while the uncertainty increases by $2\%$.
Similarly, a fit restricted to the 18 bins with $\eta(\BDT2)>0.94$ yields a change in $\mu$ of less than $0.1$, while the uncertainty increases by $3\%$. Additional fits are conducted to study the stability of the result with respect to $q^2_{\mathrm{rec}}$. In these fits, the \B background normalization is fixed to its expected value due to increased uncertainties, and the normalization of the continuum is set based on the yield observed in off-resonance data. 
The fits are separately performed for the low $q^2_{\mathrm{rec}}<4 \gevgevcccc$ and high $q^2_{\mathrm{rec}}>4 \gevgevcccc$ SR bins. The results from these fits are consistent within $1.4$ standard deviations.

The ITA method is further validated by performing a branching fraction measurement of the $B
^+\to \pi^+K^0$ decay.
This decay is reconstructed by measuring the recoil of the $\pi^+$, while the $K^0$ is not directly detected. 
In this case, the $B
^+\to \pi^+K^0$ channel exhibits a signature similar to \BKnn, with comparable selection efficiency and purity. 
The known branching fraction, measured using $\KS$ in the final state, is $(2.34\pm 0.08)\times 10^{-5}$ \cite{ParticleDataGroup:2022pth}.
With respect to the nominal \BKnn analysis, the following modifications are implemented for this validation: (i) positive pion identification is used instead of kaon identification; (ii) a bin boundary of the SR in $q^2_\mathrm{rec}$ is changed from $4$~\gevgevcccc to $2$~\gevgevcccc to increase sensitivity; (iii) the fit model uses only three sources of background (continuum, neutral \B decays, charged \B decays excluding  $B^+\to \pi^+K^0$), and the signal $B^+\to \pi^+K^0$ decays; (iv) systematic uncertainties are restricted to those originating from limited sizes of the simulated samples and global normalization uncertainties; (v) the fit is restricted to the data sample collected at the \Y4S resonance.

\begin{figure}
    \centering
    \includegraphics[width=\linewidth]{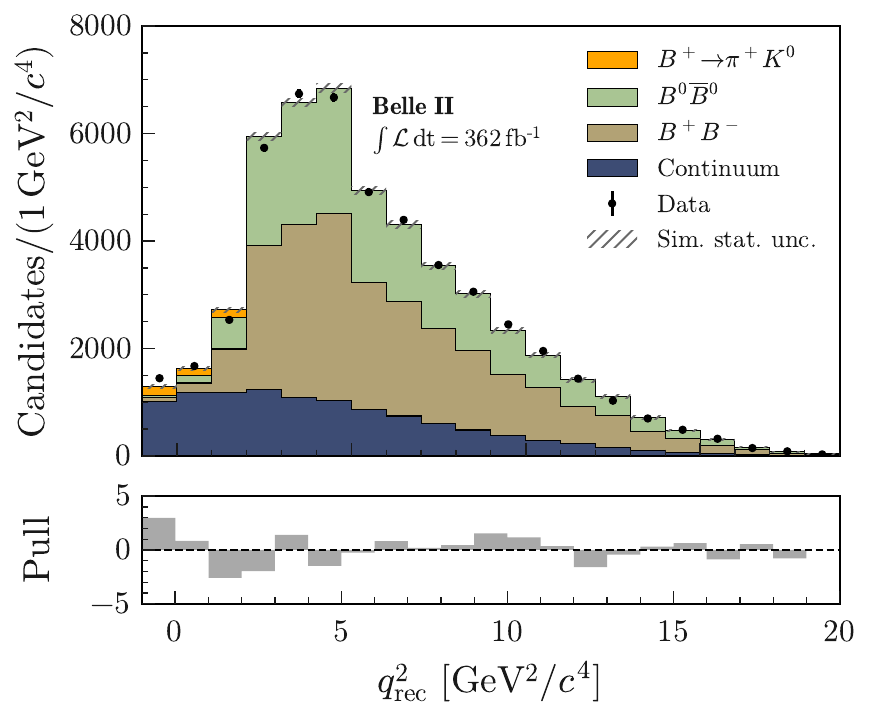}
    \caption{Distribution of  $q^2_{\mathrm{rec}}$ for ITA events in the pion-enriched sample and populating the $\eta(\BDT2) > 0.92$ bins.
    The yields of simulated background and signal components are normalized based on the fit results to determine the branching fraction of the $B^+\to \pi^+ K^0$ decay.
    The pull distribution is shown in  the bottom panel.
    }
    \label{fig:piplusfit}
\end{figure}

Based on the simulation, $80\%$ of $K^0$ within the SR are $\KL$ while the remaining $20\%$ are $\KS$.
The $B^+\to \pi^+K^0$ SR corresponds to a signal-selection efficiency of 4.4\% with 0.9\% purity, which can be compared to the 8\% and 0.9\% values for the \BKnn,  respectively.
However, the yield is almost 3 times higher, providing a sensitive test of the SR modeling. The fit quality is good,
with a $p$-value of $83\%$. 
The branching fraction of the $B^+\to \pi^+K^0$ decay is found to be $(2.5 \pm 0.5)\times 10^{-5}$, consistent with 
the known value. The distribution of $q^2_{\mathrm{rec}}$ with the background and signal components normalized using the fit result is shown in \cref{fig:piplusfit}.
 The distribution of $q^2_{\mathrm{rec}}$ for events with $\eta(\mathrm{BDT}_2) >$ 0.98 is shown in Supplemental Material~\cite{supplemental}.

\section{Combination} 
\label{sec:combination}
The consistency of the two analyses and the small size of the overlap between the HTA and ITA samples allows the combination of the results, which achieves a 10\% increase in precision over the ITA result alone.
This is done through a profile likelihood fit, incorporating correlations between common systematic uncertainties. In order to eliminate statistical correlation, common data events are excluded from the ITA dataset prior to the combination. 
Nuisance parameters corresponding to the number of $\BBbar$ events, signal form factors, and branching fractions for processes $B \to K^+\KL\KL$, $B\to D^{(**)}X$, $B\to K^+ n\bar{n}$, $D \to \KL X$, and other leading \B-meson decays are treated as fully correlated. 
To capture full correlations for the systematic uncertainties
related to the branching fractions of leading \B-meson decays between the ITA and HTA, eigendecomposition of the shared covariance matrix between ITA and HTA is performed and represented using ten nuisance parameters.

Conversely, other sources are considered uncorrelated due to their analysis-specific nature, distinct evaluation methods, or minor impact, such as PID uncertainties.

In order to ensure robustness, various scenarios are studied, including variations in which sources, such as global background normalization, are assumed to be fully correlated between the two analyses. These tests yield no substantial deviation from the default combination.

The combined result for the signal strength yields $\mu = 4.6 \pm 1.0\mathrm{(stat)} \pm 0.9\mathrm{(syst)} = 4.6 \pm 1.3 $, corresponding to a branching fraction of the $B^+\to K^+\nu\bar{\nu}$ decay of \combinationBFdetailed=\combinationBF. The significance with respect to the background-only hypothesis is found to be \combinationsigO standard deviations. 
The combined result is \combinationsigSM standard deviations above the SM expectation.

\section{Discussion} \label{sec:discussion}
\begin{figure}
    \centering
    \includegraphics[width=\linewidth]{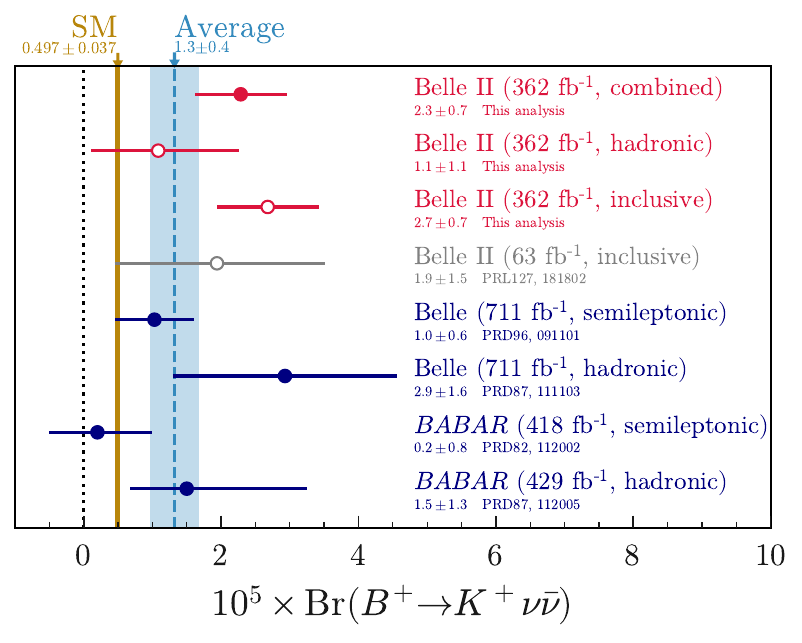}
    \caption{
    Branching-fraction values measured by Belle II, measured by previous experiments \cite{PhysRevD.82.112002,PhysRevD.87.111103,PhysRevD.87.112005,PhysRevD.96.091101,Belle-II:2021rof}, and predicted by the SM~\cite{Parrott:2022zte}.
  The Belle analyses reported upper limits; the values shown here are computed based on the quoted observed number of events, efficiency, and $f^{+-} = 0.516$. The \textit{BABAR} results are taken directly from the publications, and they use  $f^{+-}=0.5$.
  The weighted average is computed assuming symmetrized and uncorrelated uncertainties, excluding the superseded measurement of Belle II (63\invfb, inclusive) ~\cite{Belle-II:2021rof} and the uncombined results of Belle II shown as open data points.}
    \label{fig:brfr}
\end{figure}
The measured branching fraction is compared with previous measurements in \cref{fig:brfr}. The comparison is performed using branching fractions from prior measurements to assess both compatibility and relative accuracy. For \textit{BABAR}, the branching fractions are taken as given in Refs.~\cite{PhysRevD.82.112002,PhysRevD.87.112005}. Since Belle did not report branching fractions in Refs.~\cite{PhysRevD.87.111103,PhysRevD.96.091101} they are computed for this comparison based on the quoted observed number of events and efficiency taking into account statistical and systematic uncertainties. Note that \textit{BABAR} uses a different value of $f^{+-} = 0.5$ compared to the one adopted here. However, due to the large statistical uncertainties, minor differences in the correction factors have a small impact on the comparison of the results.

The ITA result is in agreement with the previous measurements obtained using hadronic and inclusive tagging methods. There are tensions of 2.3 and 1.8 standard deviations with the results obtained using semileptonic tagging by the \textit{BABAR}~\cite{PhysRevD.82.112002} and Belle~\cite{PhysRevD.96.091101} Collaborations, respectively.
The HTA result is in agreement with all measurements. The precision of the ITA measurement is comparable with the previous best results, despite being obtained with a smaller data sample. The precision of the HTA result exceeds that achieved by previous analyses using hadronic tagging. The combined Belle II result has comparable accuracy to the best single measurement, reported by Belle using semileptonic tags.

A simplified weighted average of the five independent measurements, obtained using symmetrized uncertainties (see \cref{fig:brfr}), yields a branching fraction of ${\left(1.3\pm 0.4\right)\times 10^{-5}}$, and the corresponding $\chi^2$ per degree of freedom is found to be $5.6/5$, corresponding to a $p$-value of $35\%$.

The analysis was initially performed in a manner designed to reduce experimenter's bias. The full analysis procedure was developed and finalized before determining the branching fraction from data.
However, several checks and corrections were applied after the result was obtained.
The original measurement was initially limited to the ITA and optimized through simulation using a partial data set of $189\invfb$.
In spring 2022, a fit to the data revealed a significant deviation from the expectations of the SM.  To validate the findings, the ITA was repeated using a
larger data sample while maintaining the selection criteria employed in the original measurement. As an additional consistency check, the HTA was introduced.
The new analyses underwent rigorous consistency checks before the signal strength was once again unveiled in spring 2023.
The ITA and HTA results were found to be in agreement, confirming the results of the original 2022 analysis.
Further comprehensive checks were conducted in PID sidebands, leading to changes in background modeling and an increase in systematic uncertainties.

The postunveiling changes in the ITA are corrections to the $\KL$ reconstruction efficiency in the ECL and its uncertainty, motivated by the observed excess in the pion-enriched sample (\cref{sec:klmod}); correction to the rate of $D$-meson decays involving $\KL$ and its uncertainty (\cref{sec:dkl}); and corrections to the $B^+ \to K^+\Kz \Kzb$ decay modeling and corresponding uncertainty (\cref{sec:btokpklkl}). In addition, the treatment of the reconstructed hadronic energy in the ECL was adjusted. Instead of keeping the scale at the nominal value, it is now adjusted to the preferred value while keeping the $100\%$ uncertainty (\cref{sec:neutral}). 
These modifications lead to a shift of the signal strength $\mu$ in the ITA of about $-0.5$. A mistake was found in the treatment of the $B^+\to \tau^+(\to K^+\bar{\nu})\nu$ background which was accidentally removed from the simulation. The mistake was corrected yielding a $-0.15$ change in $\mu$. Given updates of the input variables, a new training of the $\BDT2$ was performed that led to an additional $-0.5$ change in $\mu$ with a statistical uncertainty of $0.6$ estimated using simulated experiments. The modifications lead to an increase of the total uncertainty by $10\%$, driven by the uncertainty on the $B^+\to K^+\KL\KL$ branching fraction.

The HTA is based on a standard FEI data selection that is widely used within Belle II. 
During the review of another Belle II analysis~\cite{Belle-II:2023aih}, it was concluded that it is necessary to remove selection criteria on the total energy in the ECL that is poorly modeled in simulation.
The selection was removed and the BDTh was then retrained on new selected samples. This change resulted in a change of signal strength of $-2.6$. Additional HTA changes include systematic uncertainty due to the $\KL$ reconstruction efficiency in the ECL (\cref{sec:klmod}); correction to the rate of $D$-meson decays involving $\KL$ and its uncertainty
(\cref{sec:dkl}); corrections to the $B^+ \to K^+\KL \KL$ decay modeling and corresponding uncertainty (\cref{sec:btokpklkl}).
Dedicated studies were performed targeting the \Eecl variable that is correlated with the total energy in the ECL, as described in \cref{sec:neutral}, resulting in a data-driven correction and additional systematic uncertainty.
These changes resulted in a change in the signal strength of $-1.1$ with a statistical uncertainty of $1.2$, estimated using simulated experiments, which accounts for both data and simulated samples. 
The previously underestimated contributions from $\Bp \to K^+\KL\KL$ and $D\to \KL X$ background reduce the signal strength by $-0.6$. 
Taking this reduction and the estimate of the statistical uncertainty into account, the significance of the change in $\mu$ is 1.9 standard deviations.
The total uncertainty for the HTA is reduced by about $20\%$. The increase in the systematic uncertainty, also observed in ITA, is compensated by an increase in the data-sample size due to changes in the FEI selection.
\section{Summary} \label{sec:summary}
In summary, a search for the rare decay \BKnn is reported using an inclusive tagging approach with data collected by the Belle II detector at the \Y4S resonance,
corresponding to an integrated luminosity of 362\invfb. The search is validated by a well-established approach using hadronic \B tagging. The background processes are suppressed by exploiting distinct kinematic properties of the \BKnn decays in a multivariate classifier that is optimized using simulated data. The quality of the simulation is validated using several control channels. A sample-composition 
fit is used to extract the branching fraction of the \BKnn decay. 
The branching fraction obtained using the inclusive tagging is \ITABF. This measurement has a significance of \ITAsigO standard deviation with respect to the background-only hypothesis and shows a \ITAsigSM standard deviation departure from the standard model expectation.
The branching fraction obtained using the hadronic tagging is \HTABF and is consistent with the inclusive result at 1.2 standard deviations.
A combination of the inclusive and hadronic tagging results yields \combinationBF
for the \BKnn decay branching fraction, providing the first evidence of the decay with a significance of \combinationsigO standard deviations. 
The combined result shows a departure of \combinationsigSM standard deviations from the standard model expectation.

This work, based on data collected using the Belle II detector, which was built and commissioned prior to March 2019, was supported by
Higher Education and Science Committee of the Republic of Armenia Grant No.~23LCG-1C011;
Australian Research Council and Research Grants
No.~DP200101792, %
No.~DP210101900, %
No.~DP210102831, %
No.~DE220100462, %
No.~LE210100098, %
and
No.~LE230100085; %
Austrian Federal Ministry of Education, Science and Research,
Austrian Science Fund
No.~P~31361-N36
and
No.~J4625-N,
and
Horizon 2020 ERC Starting Grant No.~947006 ``InterLeptons'';
Natural Sciences and Engineering Research Council of Canada, Compute Canada and CANARIE;
National Key R\&D Program of China under Contract No.~2022YFA1601903,
National Natural Science Foundation of China and Research Grants
No.~11575017,
No.~11761141009,
No.~11705209,
No.~11975076,
No.~12135005,
No.~12150004,
No.~12161141008,
and
No.~12175041,
and Shandong Provincial Natural Science Foundation Project No.~ZR2022JQ02;
the Czech Science Foundation Grant No.~22-18469S;
European Research Council, Seventh Framework PIEF-GA-2013-622527,
Horizon 2020 ERC-Advanced Grants No.~267104 and No.~884719,
Horizon 2020 ERC-Consolidator Grant No.~819127,
Horizon 2020 Marie Sklodowska-Curie Grant Agreement No.~700525 ``NIOBE''
and
No.~101026516,
and
Horizon 2020 Marie Sklodowska-Curie RISE project JENNIFER2 Grant Agreement No.~822070 (European grants);
L'Institut National de Physique Nucl\'{e}aire et de Physique des Particules (IN2P3) du CNRS
and
L'Agence Nationale de la Recherche (ANR) under Grant No.~ANR-21-CE31-0009 (France);
BMBF, DFG, HGF, MPG, and AvH Foundation (Germany);
Department of Atomic Energy under Project Identification No.~RTI 4002,
Department of Science and Technology,
and
UPES SEED Funding Programs
No.~UPES/R\&D-SEED-INFRA/17052023/01 and
No.~UPES/R\&D-SOE/20062022/06 (India);
Israel Science Foundation Grant No.~2476/17,
U.S.-Israel Binational Science Foundation Grant No.~2016113, and
Israel Ministry of Science Grant No.~3-16543;
Istituto Nazionale di Fisica Nucleare and the Research Grants BELLE2;
Japan Society for the Promotion of Science, Grant-in-Aid for Scientific Research Grants
No.~16H03968,
No.~16H03993,
No.~16H06492,
No.~16K05323,
No.~17H01133,
No.~17H05405,
No.~18K03621,
No.~18H03710,
No.~18H05226,
No.~19H00682, %
No.~20H05850,
No.~20H05858,
No.~22H00144,
No.~22K14056,
No.~22K21347,
No.~23H05433,
No.~26220706,
and
No.~26400255,
the National Institute of Informatics, and Science Information NETwork 5 (SINET5), 
and
the Ministry of Education, Culture, Sports, Science, and Technology (MEXT) of Japan;  
National Research Foundation (NRF) of Korea Grants
No.~2016R1\-D1A1B\-02012900,
No.~2018R1\-A2B\-3003643,
No.~2018R1\-A6A1A\-06024970,
No.~2019R1\-I1A3A\-01058933,
No.~2021R1\-A6A1A\-03043957,
No.~2021R1\-F1A\-1060423,
No.~2021R1\-F1A\-1064008,
No.~2022R1\-A2C\-1003993,
and
No.~RS-2022-00197659,
Radiation Science Research Institute,
Foreign Large-Size Research Facility Application Supporting project,
the Global Science Experimental Data Hub Center of the Korea Institute of Science and Technology Information
and
KREONET/GLORIAD;
Universiti Malaya RU grant, Akademi Sains Malaysia, and Ministry of Education Malaysia;
Frontiers of Science Program Contracts
No.~FOINS-296,
No.~CB-221329,
No.~CB-236394,
No.~CB-254409,
and
No.~CB-180023, and SEP-CINVESTAV Research Grant No.~237 (Mexico);
the Polish Ministry of Science and Higher Education and the National Science Center;
the Ministry of Science and Higher Education of the Russian Federation
and
the HSE University Basic Research Program, Moscow;
University of Tabuk Research Grants
No.~S-0256-1438 and No.~S-0280-1439 (Saudi Arabia);
Slovenian Research Agency and Research Grants
No.~J1-9124
and
No.~P1-0135;
Agencia Estatal de Investigacion, Spain
Grant No.~RYC2020-029875-I
and
Generalitat Valenciana, Spain
Grant No.~CIDEGENT/2018/020;
National Science and Technology Council,
and
Ministry of Education (Taiwan);
Thailand Center of Excellence in Physics;
TUBITAK ULAKBIM (Turkey);
National Research Foundation of Ukraine, Project No.~2020.02/0257,
and
Ministry of Education and Science of Ukraine;
the U.S. National Science Foundation and Research Grants
No.~PHY-1913789 %
and
No.~PHY-2111604, %
and the U.S. Department of Energy and Research Awards
No.~DE-AC06-76RLO1830, %
No.~DE-SC0007983, %
No.~DE-SC0009824, %
No.~DE-SC0009973, %
No.~DE-SC0010007, %
No.~DE-SC0010073, %
No.~DE-SC0010118, %
No.~DE-SC0010504, %
No.~DE-SC0011784, %
No.~DE-SC0012704, %
No.~DE-SC0019230, %
No.~DE-SC0021274, %
No.~DE-SC0021616, %
No.~DE-SC0022350, %
No.~DE-SC0023470; %
and
the Vietnam Academy of Science and Technology (VAST) under Grants
No.~NVCC.05.12/22-23
and
No.~DL0000.02/24-25.

These acknowledgements are not to be interpreted as an endorsement of any statement made
by any of our institutes, funding agencies, governments, or their representatives.

We thank the SuperKEKB team for delivering high-luminosity collisions;
the KEK cryogenics group for the efficient operation of the detector solenoid magnet;
the KEK computer group and the NII for on site computing support and SINET6 network support;
and the raw-data centers at BNL, DESY, GridKa, IN2P3, INFN, and the University of Victoria for off site computing support.

\bibliography{references}

\begin{thebibliography}{51}
\bibitem [{cha()}]{chargeConj}%
  \BibitemOpen
  \href@noop {} {}\bibinfo {note} {Charge-conjugate channels are implied
  throughout this letter.}\BibitemShut {Stop}%
\end{thebibliography}

\appendix

\section{LIST OF BDT PARAMETERS AND INPUT VARIABLES}

Distributions for these variables are included in the Supplemental Material\cite{supplemental}.

\subsection{Inclusive tag analysis} \label{app:ita_vars}
Table \ref{tab:parameters} presents the parameters that are used to train the classifiers \BDT1 and \BDT2 of the ITA.
Furthermore, all input variables are listed below.
Unless otherwise specified, all variables are measured in the laboratory frame. Each variable is used in \BDT1, \BDT2, or in both BDTs as specified in parentheses.
The variable selection is done by iteratively removing variables from the training and checking the impact of their removal on the binary classification performance, measured with the area under the receiver operating characteristic curve \cite{FAWCETT2006861}.

\begin{table}[htp]
  \caption{\small Parameter values of the ITA classifier model \cite{FastBDTBelleII}.}
  \label{tab:parameters}
\begin{center}\begin{tabular}{l@{\hskip 1cm}l@{\hskip 1cm}l}
    \hline
    Parameter & Value \\
    \hline
    Number of trees & 2000 \\
    Tree depth & 2 (BDT$_1$), 3 (BDT$_2$)\\
    Shrinkage & 0.2 \\
    Sampling rate & 0.5 \\
    Number of bins & 256 \\
    \hline
  \end{tabular}\end{center}
\end{table}

For a given track, the point of closest approach (POCA) is defined as the point on the track that minimizes the distance to a line $d$ passing through the average interaction point (IP) and parallel to the $z$ axis, defined as the symmetry axis of the solenoid.
The transverse impact parameter $\dr$ is defined as this minimal distance and the longitudinal impact parameter $\dz$ is defined as the $z$ coordinate of the POCA with respect to the average interaction point \cite{BERTACCHI2021107610}. \newline

\noindent Variables related to the kaon candidate are as follows:
\begin{enumerate}[(i)]
\item radial distance between the POCA of the $K^{+}$ candidate track and the IP (BDT$_2$),%
\item cosine of the angle between the momentum line of the signal-kaon candidate and the $z$ axis (BDT$_2$). %
\end{enumerate}
Variables related to the kaon candidate do not include $q^2_{\mathrm{rec}}$, because the data are binned in this variable and in $\BDT2$ in the last stage of the analysis. \newline

\noindent Variables related to the tracks and energy deposits of\\ the rest of the event (ROE) are as follows:
\begin{enumerate}[(i)]
\item two variables corresponding to the $x$ and $z$ components of the vector from the average interaction point to the ROE vertex (BDT$_2$), %
\item $p$-value of the ROE vertex fit (BDT$_2$), %
\item variance of the transverse momentum of the ROE tracks (BDT$_2$), %
\item polar angle of the ROE momentum (BDT$_1$, BDT$_2$), %
\item magnitude of the ROE momentum (BDT$_1$, BDT$_2$), %
\item a modified Fox-Wolfram moment of the ``oo'' type (see Ref.~\cite{Bevan:2014iga}), i.e.,\ ROE-ROE, calculated in the c.m.\ frame (BDT$_1$, BDT$_2$),%
\item difference between the ROE energy in the c.m.\ frame and the energy of one beam in the c.m.\ frame ($\sqrt{s}/2$)\ (BDT$_1$, BDT$_2$).  %
\end{enumerate}

\noindent Variables related to the entire event are as follows:
\begin{enumerate}[(i)]
\item number of \epm and {\ensuremath{\mu^\pm}\xspace} candidates (BDT$_2$), %
\item number of photon candidates, number of charged particle candidates (BDT$_2$), %
\item square of charged particles in the event (BDT$_2$),%
\item cosine of the polar angle of the thrust axis in the c.m.\ frame (BDT$_1$, BDT$_2$), %
\item harmonic moments with respect to the thrust axis in the c.m.\ frame~\cite{Fox:1978vw} (BDT$_1$, BDT$_2$), %
\item modified Fox-Wolfram moments calculated in the c.m.\  frame~\cite{PhysRevLett.91.261801} (BDT$_1$, BDT$_2$), %
\item polar angle of the missing three-momentum in the c.m.\ frame (BDT$_2$), %
\item square of the missing invariant mass (BDT$_2$),%
\item event sphericity in the c.m.\ frame~\cite{Bevan:2014iga} (BDT$_2$), %
\item normalized Fox-Wolfram moments in the c.m.\ frame~\cite{Fox:1978vw} (BDT$_1$, BDT$_2$), %
\item cosine of the angle between the momentum of the signal-kaon and the ROE thrust axis in the c.m.\ frame~(BDT$_1$, BDT$_2$),%
\item radial and longitudinal distance between the POCA of the $K^+$ candidate track and the tag vertex (BDT$_2$). %
\end{enumerate}

\noindent Variables related to the $\Dz/\Dp$ suppression\\ $\Dz$ candidates are obtained by fitting the kaon candidate track and each track of opposite charge in the ROE to a common vertex; $\Dp$ candidates are obtained by fitting the kaon candidate track and two ROE tracks of appropriate charges. In both cases, we choose the candidate having the best vertex fit quality. The related variables are as follows:
\begin{enumerate}[(i)]
\item radial distance between the chosen $D^+$ candidate vertex and the IP (BDT$_2$), %
\item $\chi^2$ of the chosen $\Dz$ candidate vertex fit and the best $\Dp$ candidate vertex fit (BDT$_2$), %
\item mass of the chosen $\Dz$ candidate (BDT$_2$), %
\item median $p$-value of the vertex fits of the $\Dz$ candidates (BDT$_2$). %
\end{enumerate}

\subsection{Hadronic tag analysis} \label{app:hta_vars}
BDTh parameters, reported in~\cref{tab:parameters_hta}, are optimized based on a grid search in the parameter space.
The following 12 variables
are used as input:
\begin{enumerate}[(i)]
\item sum of extra-photon energy deposits in ECL,
\item number of extra tracks,
\item sum of the  missing energy and absolute missing three-momentum vector,
\item azimuthal angle between the signal kaon and the missing-momentum vector,
\item cosine of the angle between the momentum direction of the signal-kaon candidate, and the thrust axis of the particles comprising the $B_{\rm tag}$, the extra tracks, and the extra photons,

\item modified Fox-Wolfram moments $H^{\rm so}_{22}$, $H^{\rm so}_{02}$, $H^{\rm oo}_{0}$, %
\item invariant mass of of the four-momentum difference between the two colliding beams and the signal kaon,
\item signal probability for the $B_{\rm{tag}}$ returned by the FEI algorithm,
\item $p$-value of the vertex fit of the signal kaon and one or two tracks in the event to reject fake kaons coming from \Dz or \Dp decays.
\end{enumerate}

\begin{table}[htp]
  \caption{\small Parameter values of the HTA classifier model \cite{FastBDTBelleII}.}
  \label{tab:parameters_hta}
\begin{center}\begin{tabular}{l@{\hskip 1cm}l@{\hskip 1cm}l}
    \hline
    Parameter & Value \\
    \hline
    Number of trees & 1300 \\
    Tree depth & 3 \\
    Shrinkage & 0.03\\
    Sampling rate & 0.8\\
    Number of equal-frequency bins & 256\\
    \hline
  \end{tabular}\end{center}
\end{table}

\section{STUDIES OF THE PION-ENRICHED SAMPLE IN THE ITA}
\label{sec:pionfits}

The pion-enriched sample in the ITA is chosen by applying the selection criteria for the SR of the nominal analysis, with the PID requirement for the signal candidate suitably adjusted. The simulated samples are corrected for known data-to-simulation discrepancies for the pion identification and kaon-to-pion as well as lepton-to-pion fake rates. The modeling of the continuum background is improved using BDT$_c$, which is taken from the nominal analysis without retraining. The normalization of the continuum background is determined using the off-resonance data sample and is found to be $1.30 \pm 0.03$ compared to the expectations from simulation. Simulated $B$-meson samples are divided into two subsamples based on the presence of a $D$ meson in the decay chain associated with the signal pion, where the $D$ meson's decay products include a \KL meson. They are normalized using the number of the $\BBbar$ pairs collected by Belle II. 

Figure~\ref{fig:pionSBfits} shows the $q^2_{\mathrm{rec}}$ distribution in data and simulation. 
A discrepancy in the data-to-simulation ratio is observed when the normalization factors described above are used, as can be seen from the left panel of the figure. The $q^2_{\mathrm{rec}}$ variable corresponds to the invariant mass squared of the system recoiling against the candidate pion in the $B$ decay. 
The data-to-simulation ratio is below unity below the threshold of the $D$-meson mass and above unity beyond the threshold.
For $q^2_{\mathrm{rec}}<0$, which is dominated by continuum background, the data are below the simulation, suggesting that the scale factor determined based on the off-resonance sample is overestimated. A binned sample-composition fit is performed to this $q^2_{\mathrm{rec}}$ distribution to determine the normalization of the three components. 
In this fit, the sample of $B$-meson decays containing $D$-mesons with $\KL$ in the decay chain is allowed to vary without constraints, while the complementary $B$-meson decay sample is constrained by the uncertainty associated with the number of $\BBbar$ pairs. The uncertainty on the continuum normalization is taken to be $50\%$, as in the nominal fit. The result of this fit is shown in the right panel of the figure, and shows a much improved data-to-simulation ratio. The shift in the normalization factor for the sample including $D \to \KL X$ decays is found to be $(+30\pm 2)\%$: this shift is used in the nominal analysis. The normalization factor for the sample without $D$ mesons involving $\KL$ is consistent with the expectations to within $1.2$ standard deviations, while the normalization for the continuum background is reduced by $(19 \pm 2)\%$ and is therefore closer to the expectation from the simulation.
The difference in the continuum normalization based on the off- and on-resonance data from this fit motivates the systematic uncertainty on the off-resonance sample normalization applied in the nominal analysis.
\newline

\section{RECONSTRUCTION OF $\boldsymbol{B^+\rightarrow K^+\KS\KS}$ AND $\boldsymbol{B^0 \rightarrow \KS K^+K^-}$ CONTROL DECAY CHANNELS}
\label{app:klkl}

Reconstruction of $B^+\to K^+\KS\KS$ and $B^0 \to \KS K^+K^-$ decay channels utilizes the $K^+$ selection adapted for the main analysis.
The $\KS$ candidates are reconstructed using the $\KS \to \pi^+\pi^-$ decay mode, and a multivariate classifier is employed to enhance sample purity. The signal region is defined by requiring $M_{\rm{bc}}>5.27 \gevcc$ and $|\Delta E|<0.2\gev$. The average multiplicity of candidates is about 1.5 for this selection.
In cases where multiple candidates are found, a single candidate per event is randomly selected. For the less pure $B^+ \to K^+ \KS\KS$ decay, the input variables and classifier parameters are set to the same values as those used for \BDT2 in the nominal measurement. For the $B^0\to \KS K^+K^-$ decay, \BDT1 variables and parameters are utilized.
Less stringent criteria on the output of the classifiers are applied and checked using simulated signal samples to ensure approximately constant efficiency as a function of $M(\KS\KS)$ and $M(K^+K^-)$ for each decay, respectively.

The signal is extracted by an unbinned maximum likelihood fit to the $\Delta E$ distribution. For the $B^+ \to K^+ \KS\KS$ decay, a Gaussian distribution is used to model the signal.
For the $B^0 \to \KS K^+K^-$ decay, the signal is modeled by a sum of two Crystal Ball functions~\cite{Skwarnicki:1986xj} plus a sum of two Gaussian distributions with the shape of the model fixed by the simulation and the total width allowed to vary in the fit to data. The background is modeled by a second-order Chebyshev polynomial in both cases.

\end{document}